\documentclass[aps,prx,superscriptaddress,twocolumn,nofootonbib,floatfix,longbibliography]{revtex4-2} 

\usepackage[final]{graphicx}    
\usepackage{soul}
\usepackage{enumitem}
\usepackage[utf8]{inputenc}
\usepackage{natbib}
\usepackage{xcolor}
\usepackage{amsfonts}
\usepackage{amsmath}
\usepackage{hyperref}
\hypersetup{colorlinks=true}

\newcommand{\mbf}[1]{\mathbf{#1}}

\makeatletter
\newcommand\footnoteref[1]{\protected@xdef\@thefnmark{\ref{#1}}\@footnotemark}
\makeatother

\begin{document}

\title{Low-depth Clifford circuits approximately solve MaxCut}

\author{Manuel H. Muñoz-Arias}
\email{munm2002@usherbrooke.ca}
\affiliation{Institut Quantique and Département de Physique, Université de Sherbrooke, Sherbrooke, QC J1K 2R1, Canada}
\author{Stefanos Kourtis}
\affiliation{Institut Quantique and Département de Physique, Université de Sherbrooke, Sherbrooke, QC J1K 2R1, Canada}
\author{Alexandre Blais}
\affiliation{Institut Quantique and Département de Physique, Université de Sherbrooke, Sherbrooke, QC J1K 2R1, Canada}
\affiliation{Canadian Institute for Advanced Research, Toronto, ON M5G 1M1, Canada}

\begin{abstract}
We introduce a quantum-inspired approximation algorithm for MaxCut based on low-depth Clifford circuits. We start by showing that the solution unitaries found by the adaptive quantum approximation optimization algorithm (ADAPT-QAOA) for the MaxCut problem on weighted fully connected graphs are (almost) Clifford circuits. Motivated by this observation, we devise an approximation algorithm for MaxCut, \emph{ADAPT-Clifford}, that searches through the Clifford manifold by combining a minimal set of generating elements of the Clifford group. Our algorithm finds an approximate solution of MaxCut on an $N$-vertex graph by building a depth $O(N)$ Clifford circuit. The algorithm has runtime complexity $O(N^2)$ and $O(N^3)$ for sparse and dense graphs, respectively, and space complexity $O(N^2)$, with improved solution quality achieved at the expense of more demanding runtimes. We implement ADAPT-Clifford and characterize its performance on graphs with positive and signed weights. The case of signed weights is illustrated with the paradigmatic Sherrington-Kirkpatrick model, for which our algorithm finds solutions with ground-state mean energy density corresponding to $\sim94\%$ of the Parisi value in the thermodynamic limit. The case of positive weights is investigated by comparing the cut found by ADAPT-Clifford with the cut found with the Goemans-Williamson (GW) algorithm. For both sparse and dense instances we provide copious evidence that, up to hundreds of nodes, ADAPT-Clifford finds cuts of lower energy than GW.

\end{abstract}

\date{\today}
\maketitle

\section{Introduction}
\label{sec:intro} 

Near-term quantum processors have found a niche in the hybrid quantum-classical model of computation with Variational Quantum Algorithms (VQAs)~\cite{McClean2016,Cerezo2021,Bharti2022}. These algorithms perform \emph{classical} optimization of a problem-specific objective function that is evaluated by measuring the output of a parametrized \emph{quantum} circuit. Through variational search over circuit parameters, a VQA thus seeks a solution circuit that transforms a simple input state in the Hilbert space of problem variables to a superposition of approximate solutions, i.e., configurations that yield near-optimal values of the objective function. In attempts to determine whether they can lead to a speedup over classical algorithms for any useful task, VQAs have been applied to a variety of combinatorial optimization problems. 

Quantum approximate optimization algorithms (QAOAs)~\cite{Farhi2014,Hadfield2019} have been in a constant tug of war with classical solvers, with initial indications of putative quantum speedups~\cite{Crooks2018,Golden2022,Boulebnane2022,Carlson2023}, followed by experimental claims~\cite{Ebadi2022} and rebuttals~\cite{Andrist2023}, and then further proposals for possible quantum speedup~\cite{Shaydulin2023}. A byproduct of this large effort has been the definition and construction of quantum-inspired algorithms~\cite{Gilyen2018,Tang2019,Gilyen2022,Arrazola2020,Tene2023,Misra2023}. These are ``dequantized'' classical versions of quantum or hybrid quantum-classical algorithms that unveil and exploit previously unrecognized properties or structures in a problem, leading to solution strategies that outperform the best known classical algorithm.

In this work, we introduce a quantum-inspired approximation algorithm for the MaxCut problem. The algorithm, which we dub ADAPT-Clifford, is motivated by the observation that the solution circuits found by an adaptive QAOA variant for MaxCut on weighted complete graphs are (almost) Clifford circuits, a well-known restricted class of quantum circuits that are easy to simulate classically~\cite{Gottesman1996,Gottesman1997,Gottesman1998,Aronson2004,Gidney2021stimfaststabilizer}. ADAPT-Clifford builds an entangled state with a number of unitary operations that is equal to the number of nodes $N$, adding at every step a known two-qubit gate to the circuit. The algorithm is polynomial both in time and space, with worst case runtime complexity $O(N^4)$ and space complexity $O(N^2)$. We characterize the performance of the algorithm in several families of graphs. For graph sizes up to $N=30$ nodes we report the exact approximation ratios. For larger problem sizes up to hundreds of nodes and depending on graph family, we assess the performance by either direct comparison with the solution found by the best classical algorithm for MaxCut~\cite{Goemans1995,Goemans1995a}, or by comparing with the known value of the mean energy density in the thermodynamic limit.

The rest of this manuscript is organized as follows. In Sec.~\ref{sec:background} we present a short summary of the MaxCut problem, quantum approximate optimization and its adaptive variant, Clifford circuits, and introduce the tools we use to characterize the structure of solution circuits. In Sec.~\ref{sec:algo_origin} we present the origin of the algorithm by analyzing the structure of solution circuits to MaxCut on weighted complete graphs obtained with QAOA and ADAPT-QAOA. In Sec.~\ref{sec:clifford_algo} we present the ADAPT-Clifford algorithm, discuss its details, and analyze its runtime and space complexities in different types of graphs. In Sec.~\ref{sec:performance_on_complete} we present numerical results of the algorithm performance on weighted complete graphs with positive and signed weights, using the Sherrington-Kirkpatrick model as a specific example of the latter. In Sec.~\ref{sec:limitations} we explore the performance of the algorithm beyond complete graphs, including weighted and unweighted $K$-regular graphs and Erdös-Rényi graphs. Finally in Sec.~\ref{sec:outlook} we conclude with a discussion of our results in the context of near-term quantum optimization algorithms and present an outlook of future work. 

\section{Background and methods}
\label{sec:background}

\subsection{The MaxCut problem} 
Given a graph $\mathcal{G}=(\mathcal{V},\mathcal{E})$, where $\mathcal{V}$ is the vertex set and $\mathcal{E}\subseteq\mathcal{V}^2$ is the set of edges ($\mathcal{E}=\mathcal{V}^2$ is a complete graph), and edge weights $\omega_{i,j}\in\mathbb{R}$ for $(i,j)\in\mathcal{E}$, the MaxCut problem asks to partition $\mathcal{V}$ into two complementary subsets $\mathcal{A},\overline{\mathcal{A}}\subset\mathcal{V}$, such that the total weight of the edges between $\mathcal{A}$ and $\overline{\mathcal{A}}$ is maximized. We use binary variables $z_i\in\{0,1\}$, $i\in\mathcal{V}$ to help us identify each subset, so that $z_i=1$ if vertex $i\in\mathcal{A}$ and $z_i=0$ if $i\in\overline{\mathcal{A}}$. The maximal cut can then be formally expressed as the assignment $\mbf{z}'$ that maximizes the cost function 
\begin{equation}
\label{eqn:classical_cost}
C(\mathbf{z}) = \sum_{(i,j)\in\mathcal{E}}\omega_{i,j}z_i(1-z_j),
\end{equation}
where $\mathbf{z}=z_1...z_N$ is a $N$-bit binary string and $\omega_{i,j} = \omega_{j,i}$ $\forall(i,j)\in\mathcal{E}$.

The MaxCut problem on general graphs is known to be NP-hard~\cite{Karp1972}\footnote{There is no polynomial time algorithm to solve the problem exactly in the worst case. It has also shown to be APX-hard~\cite{Papadimitriou1988} no polynomial time approximation scheme exists unless P=NP.}. However, MaxCut can be solved in polynomial time in some special cases, such as graphs without long odd cycles~\cite{Grotschel1984}, weakly bipartite graphs~\cite{Grotschel1981}, planar graphs both weighted~\cite{Shih1990,Liers2012} and unweighted~\cite{Hadlock1975,Orlova1972}, $1$-planar graphs~\cite{Dahn2018}, and graphs with $k$ crossings~\cite{Chiami2020}. Finally, when all the edge weights are negative, MaxCut becomes a equivalent to MinCut and admits a polynomial time algorithm~\cite{Stoer1997}.

Beyond the special cases mentioned above and due to the difficulty of solving the problem exactly, one often aims instead to find approximation algorithms that yield reasonably good solutions in polynomial time for all problem instances. That is, we search for an algorithm that outputs an assignment $\mathbf{z}^*$, such that the approximation ratio 
\begin{equation}
\label{eqn:classical_ratio}
\alpha(\mathbf{z}^*) = \frac{C(\mathbf{z^*})}{\underset{\mathbf{z}}{{\rm max}}[C(\mathbf{z})]},
\end{equation}
equals some desired value, ideally as close to $1$ as possible, on all instances of MaxCut. However, in some cases the gap between approximate an optimal solutions cannot be reduced arbitrarily in polynomial time~\cite{Khot2010}, a phenomenon known as hardness of approximation. For MaxCut on general graphs, the best known approximation algorithm is that of Goemans and Williamson (GW)~\cite{Goemans1995}, which has a performance guarantee (worst case) of $\alpha \simeq 0.878$~\cite{Goemans1995,Karloff1999,Goemans1995a}~\footnote{We point out that there exist several heuristic algorithms which perform well, see Ref.~\cite{Dunning2018} for a comparison of several ones. However at the moment these have not known formal guarantees.}. Below we will present extensive performance comparisons between ADAPT-Clifford and the GW algorithm, in order to be as self contained as possible we review the details of the GW algorithm in App.~\ref{app:gw}.

One might hope to achieve better approximation ratios by focusing on specific families of graphs. For unweighted graphs Ref.~\cite{Hastad2001} showed that finding an algorithm yielding an approximation ratio better than $16/17$ is NP-hard. Nearly optimal algorithms both for cubic graphs~\cite{Halperin2004}, guaranteeing $\alpha = 0.9326$, and for $K$-regular graphs of large degree~\cite{El2021b}, are known. Another interesting example is the case of dense graphs, i.e., graphs with $O(N^2)$ edges. Polynomial time approximation schemes (PTAS) are known for both unweighted~\cite{de_la_vega1996,Arora1999} and weighted~\cite{de_la_vega2000} graphs, although for the latter there is only an existence result. A PTAS guarantees an approximate solution whose cost is $1-\epsilon$ away from the optimal. Although these schemes have a provable polynomial runtime in $N$, it might not be polynomial in $\epsilon$, see for example Ref.~\cite{Arora1999}. We will come back to this point in Sec.~\ref{sec:outlook}. 

In quantum approximate optimization, the objective function of a combinatorial optimization problem defined on binary variables $z_i$, such as MaxCut, is expressed as an Ising Hamiltonian through the mapping $\sigma_i = 2z_i - 1$, with connectivity dictated by the graph $\mathcal{G}$~\cite{Lucas2014}. That is, the entries $\omega_{i,j}$ of the adjacency matrix reflect the coupling between the $i$-th and $j$-th spin. In this setting the optimum $\mbf{z}'$ is encoded as the ground state of the Ising Hamiltonian. The Ising Hamiltonian is then promoted to a Hamiltonian operator via the identification $\sigma_i\to Z_i$, with $Z_i$ a Pauli-$z$ operator acting on the qubit that corresponds to the $i$-th spin. For the MaxCut problem, the corresponding Hamiltonian is 

\begin{equation}
\label{eqn:cost_hamil}
H_{\rm C} = \frac{1}{2}\sum_{i<j}\omega_{i,j}Z_iZ_j.
\end{equation}
In writing Eq.~(\ref{eqn:cost_hamil}) we have dropped a constant factor equal to $\sum_{i<j}\frac{\omega_{i,j}}{2}$ and added a minus sign to turn the maximization problem defined by Eq.~(\ref{eqn:classical_cost}) into a minimization one. 

In analogy with classical approximation algorithms, quantum approximate optimization yields approximate solutions in the form of a state $|\phi\rangle$, whose energy expectation is as close as possible to the ground-state energy of the Ising Hamiltonian. Thus, the approximation ratio, Eq.~(\ref{eqn:classical_ratio}), takes the form 
\begin{equation}
\label{eqn:aratio}
\alpha = \frac{\langle \phi|H_{\rm C}|\phi\rangle}{E^{\rm C}_{\rm min}},
\end{equation}
where $E^{\rm C}_{\rm min}$ is the smallest eigenvalue of $H_{\rm C}$. To achieve advantageous performance, a quantum algorithm must produce an approximate solution with a desired $\alpha$ faster than any classical algorithm.

\subsection{The quantum approximate optimization algorithm and its adaptive variant}
The Quantum Approximate Optimization Algorithm (QAOA) is a type of variational algorithm~\cite{Cerezo2021} that aims to solve combinatorial optimization problems~\cite{Farhi2014}. It is defined by a parametrized quantum circuit with a periodic structure. Each layer of the circuit is given by a product of two unitaries, time evolution under $H_{\rm C}$, followed by time evolution under a mixer Hamiltonian

\begin{equation}
 H_{\rm M}=\sum_{j=1}^N X_j, 
\end{equation}
where $X_j$ is a Pauli-$x$ on the $j$-th qubit. For $p$ layers QAOA prepares the state
\begin{equation}
\label{eqn:qaoa_state}
|\psi(\boldsymbol{\gamma}, \boldsymbol{\beta})\rangle_p = \left[\prod_{l=1}^p e^{-i\beta_l H_{\rm M}} e^{-i \gamma_l H_{\rm C}} \right] \mathrm{H}^{\otimes N} |0\rangle^{\otimes N},
\end{equation}
where $\boldsymbol{\gamma} = \gamma_1,...,\gamma_p$, $\boldsymbol{\beta} = \beta_1,...,\beta_p$, and $\mathrm{H}$ is the Hadamard gate. In order to find approximate solutions the set of $2p$ parameters is optimized so as to minimize $\langle \psi(\boldsymbol{\gamma}, \boldsymbol{\beta})|H_{\rm C}|\psi(\boldsymbol{\gamma}, \boldsymbol{\beta})\rangle_p$. After executing the circuit with optimized parameters a measurement in the computational basis returns a candidate solution in the form of a bit string $\mathbf{z}^*$. Ideally, one would sample with high probability a good approximate solution. We will denote the optimal parameters found by numerical experiments as $(\boldsymbol{\gamma}^*, \boldsymbol{\beta}^*)$, and the associated solution unitary $U(\boldsymbol{\gamma}^*, \boldsymbol{\beta}^*) = \prod_{l=1}^p e^{-i\beta^*_l H_{\rm M}} e^{-i \gamma^*_l H_{\rm C}}$. 

Not much is known regarding performance guarantees and hardness of approximation for QAOA. The case of constant $p$ has so far been the main focus, as it is the regime of interest for current quantum devices~\cite{Preskill2018}. Ref.~\cite{Wurtz2021} gives evidence for a possible quantum advantage for MaxCut on $3$-regular graphs with shallow QAOA. Refs.~\cite{Valle2023,Lotshaw2023} provide evidence that both $p=1$ and large-depth QAOA output states with bit string probabilities following Boltzman distributions, rendering sampling classically hard. At the same time, it is known that constant $p$ QAOA is bounded away from optimality in sparse graphs~\cite{Chou2021,Farhi2020a,Farhi2020b}, as well as in some dense problems where the overlap gap property~\footnote{The overlap gap property (OGP) is a property of the geometry of the space of solutions which (very) roughly speaking states that nearly optimal solutions are either close or far apart from each other. The OGP is currently understood as an algorithmic barrier as it implies a large family of algorithms have their best solutions bounded away from optimality. For an extended discussion see Ref.~\cite{Gamarnik2022}.} is known to exist~\cite{Basso2022a}. These results were recently extended to the case of $p\sim\log(N)$~\cite{Chen2023}. However these results do not apply directly to the case of $p\sim{\rm poly}(N)$. As a consequence there are no conclusive results on the runtime required for $p\sim{\rm poly}(N)$ QAOA to reach a given approximation ratio, with only loose lower bounds appearing recently~\cite{Benchasattabuse2023}. Most studies of QAOA so far have been numerical experiments on different families of problem instances, two examples are Erdos-Renyi graphs~\cite{Crooks2018} and $3$- and $4$-regular graphs (weighted and unweighted)~\cite{Zhou2020}. Importantly, any indication of a putative advantage in this type of studies, has been inconclusive due to the small problem sizes accessible to either quantum implementations or classical simulation~\footnote{With further putative evidence coming from numerical experiments relying on techniques to bypass the classical optimization of the variational parameters~\cite{Zhou2020,Farhi2022,Shaydulin2023,Brandao2018} or to steer the optimization process in a more efficient and effective manner~\cite{Amaro2022}.}.

 
To alleviate some of the roadblocks explored above, variants to the original QAOA ansatz have been developed, see Ref.~\cite{Blekos2023} for a review. Of interest to us here is the ADAPT~\cite{Zhu2022} variant, which was proposed as a way to find ansätze which are tailored to the specifics of the problem under consideration. ADAPT-QAOA is an iterative variational algorithm which replaces the fixed mixer Hamiltonian in Eq.~(\ref{eqn:qaoa_state}), by a suitably chosen one, $A_l$, at each layer $l\le p$. Thus, $p$-layer ADAPT-QAOA prepares the state 
\begin{equation}
\label{eqn:adapt_state}
|\psi(\gamma, \beta)\rangle_p^{\rm ADAPT} = \left[\prod_{l=1}^p e^{-i\beta_l A_l } e^{-i \gamma_l H_{\rm C}} \right] \mathrm{H}^{\otimes N} |0\rangle^{\otimes N}.
\end{equation}
The $l$-th mixer Hamiltonian is chosen as the one which maximizes the energy gradient, that is, 
\begin{equation}
\label{eqn:adapt_mixer}
A_l = \underset{A_s \in \mathrm{P}_{\rm OP}}{\rm max}\left[-i\langle\psi_{l-1}|e^{i\gamma_l H_{\rm C}}[H_{\rm C}, \hat{A}_s]e^{-i\gamma_l H_{\rm C}}|\psi_{l-1}\rangle \right],
\end{equation}
where the new variational parameter $\gamma_l$ is set to a predefined small positive value $\gamma_0\sim0$~\cite{Zhu2022}, $\mathrm{P}_{\rm OP}$ is an operator pool, and $|\psi_{l-1}\rangle$ is the state resulting from the application of the ADAPT-QAOA solution circuit with only $l-1$ layers. The choice of pool is not unique, with different pools being advantageous in different situations~\cite{Chen2022,Feniou2023}. Below we restrict ourselves to the pool
\begin{equation}
\label{eqn:operator_pool}
\begin{split}
\mathrm{P}_{\rm OP} =& \{ \sum_i X_i, \sum_i Y_i \} \cup \{X_j, Y_j\}_{j=1,...,N} \\ 
&\cup \{X_j X_k, Y_j Y_k, Y_j Z_k, Z_j Y_k\}_{j,k=1,...,N, j\ne k},
\end{split}
\end{equation}
which is sufficient for our purposes.

In contrast to QAOA, ADAPT-QAOA grows the circuit layer by layer, until the desired number $p$. As such, we begin with a single layer, find the corresponding mixer according to Eq.~(\ref{eqn:adapt_mixer}), then optimize to find the best parameters. We then add a second layer, find the corresponding mixer according to Eq.~(\ref{eqn:adapt_mixer}), initialize the new pair of parameters to zero~\footnote{In practice the structure of the cost landscape forces us to initialize $\gamma_l$ to a small positive value, which we fix to be $\gamma_l=0.01$ see Ref.~\cite{Zhu2022} for details.}, the rest of the parameters to the best values already found, and optimize all of them. This procedure is repeated until $p$ layers are added. For a fair comparison between QAOA and ADAPT-QAOA in our numerical simulations we construct the QAOA solution circuit following the same iterative strategy, but with a fixed mixer.

\subsection{Clifford circuits and their efficient simulation}
\label{subsec:clifford_Stuff}
In this subsection we review some concepts of the stabilizer formalism which will be used later in the manuscript. For a general presentation see Ref.~\cite{NielsenChuangBook}. 

The single qubit Pauli group is given by the operators $\{\mathbb{I},X,Y,Z\}$ together with multiplicative factors $\pm1,\pm i$. The $N$ qubit Pauli group $\mathcal{\tilde{P}}_N$ is given by all the $N$-tensor products of these operators together with multiplicative factors. Given a pure state on $N$ qubits $|\psi\rangle$, we say $\tilde{P}_i\in\mathcal{\tilde{P}}_N$ stabilizes $|\psi\rangle$ if the state is an eigenvector of $\tilde{P}_i$ with eigenvalue $+1$: $\tilde{P}_i|\psi\rangle = |\psi\rangle$. A $n$-qubit pure state is a stabilizer state if it can be completely specified, up to a global phase, by its $N$ stabilizers.

Quantum circuits which map stabilizer states to stabilizer states define a large class of nontrivial quantum circuits ---stabilizer circuits--- which can be simulated in polynomial time on a classical computer~\cite{Aronson2004,Gidney2021stimfaststabilizer}. This is the content of the celebrated Gottesman-Knill theorem~\cite{Gottesman1997,Gottesman1998}. These quantum circuits can be completely written in terms of controlled-NOT, Hadamard, and phase gates, and single qubit measurements. Importantly, the efficient classical simulability does not imply these circuits are not interesting. On the contrary, they have extensive applications in quantum information science, for instance encoding and decoding in quantum error correction~\cite{Bennett1996,Calderbank1997,Gottesman1996,Gottesman1998}, dense quantum coding~\cite{Bennett1992}, quantum teleportation~\cite{Bennett1993}, quantum simulation~\cite{Beguvsic2023,Beguvsic2023a}, proof of principle of quantum advantage with nonlocal games~\cite{Daniel2021,Daniel2022}, as well as in quantum many-body physics~\cite{Kitaev1997,Li2018,Chan2019,Cote2022,Kelly2022}. 

In absence of measurements, stabilizer circuits are referred to as Clifford circuits or Clifford unitaries. They form a group $\mathcal{C}$, defined as the unitaries which normalize the Pauli group, that is, the unitaries which map Pauli operators to Pauli operators. Following from the Gottesman-Knill theorem, this group has three generators, the controlled-NOT, Hadamard, and phase gates. Naturally the Pauli operators are elements of the group, as they are generated by Hadamard and phase. 

Both the QAOA and ADAPT-QAOA ansätze are defined as products of unitaries generated by Pauli strings. When are unitary transformations generated by Pauli strings Clifford unitaries? To answer this question, take $\tilde{P}_i,\tilde{P}_j\in \mathcal{\tilde{P}}$, two distinct Pauli strings that either commute or anticommute by definition. Further consider the unitary $W(\theta)=e^{-i\theta \tilde{P}_j}$, then $W^\dagger \tilde{P}_i W = \tilde{P}_i$ if $[\tilde{P}_i,\tilde{P}_j]=0$, and $W^\dagger \tilde{P}_i W = i\tilde{P}_i\tilde{P}_j$ if $\{\tilde{P}_i,\tilde{P}_j\} = 0$ and $\theta = \pm m\frac{\pi}{4}$ with $m\in\mathbb{N}$. We thus see that when a quantum circuit is composed of products of unitaries generated by Pauli strings that do not necessarily commute, it is a Clifford circuit if an only if the parameters of these transformations are integer multiples of $\pm\pi/4$. Therefore, if QAOA solution circuits $U(\boldsymbol{\gamma}^*, \boldsymbol{\beta}^*)$ are to be Clifford, then the circuit parameters $\boldsymbol{\gamma}^*$ and $\boldsymbol{\beta}^*$ must be integer multiples of $\pm\pi/4$.

\subsection{Characterizing the structure of solution unitaries}
Here we introduce the tools we use in the next section to characterize the structure of QAOA solution circuits $U(\boldsymbol{\gamma}^*, \boldsymbol{\beta}^*)$. Consider the Hilbert space $\mathcal{H}$ of $N$ qubits with dimension $d = 2^N$, and define the $N$-qubit Pauli basis as $\mathcal{P}_N = \mathcal{\tilde{P}}_N/\langle\pm i \mathbb{I}\rangle$, the quotient group containing, $D=4^N-1$, Pauli strings with all multiplicative factors equal to $+1$. Furthermore any pair of Pauli strings obey ${\rm Tr}[P_i P_j] = d\delta_{ij}$. Therefore, $\mathcal{P}_N$ defines a basis for all Hermitian operators in $\mathcal{H}$.

Consider some Hermitian operator $O$ acting on $\mathcal{H}$. If $O$ evolves under some unitary transformation $V$, we write 
\begin{equation}
\label{eqn:evolved_ope}
O' = V^\dagger O V = \sum_{j=1}^{D} f[P_j;O'] P_j,
\end{equation}
with $P_j\in\mathcal{P}_N$. Noticing that $\sum_j |f[P_j;O']|^2 = {\rm Tr}[O'^2] = {\rm Tr}[O^2]$, we define 
\begin{equation}
\label{eqn:operator_distri}
p_j(O;V) = \frac{1}{{\rm Tr}[O^2]}|f[P_j;O']|^2.
\end{equation}
It is easy to see that $\sum_j p_j = 1$. Eq.~(\ref{eqn:operator_distri}) thus denotes the probability of finding $O'$ to be the $j$-th Pauli string $P_j$. In the case of $O=P_l$, the normalization factor in Eq.~(\ref{eqn:operator_distri}) is $\sum_j |f[P_j;O']|^2 = d$.

We analyze the transformation $V$ as an ``input-output'' channel, with $O$ the input and $O'$ the output, and are interested in characterizing the locality, in the Pauli basis, of the output. This can be inferred from the  the localization properties of $p_j(O;V)$, which we investigate with the Second Renyi entropy, see Ref.~\cite{Renyi1961} and Sec. 2.7 of Ref.~\cite{Bengtsson2017}
\begin{equation}
\label{eqn:single_site_renyi}
\mathcal{S}\left(O;V\right) = -\log\left(\sum_{j=0}^{4^N}\frac{|f[P_j;O']|^4}{d^2}\right).
\end{equation}
Eq.~(\ref{eqn:single_site_renyi}) has a resemblance to the stabilizer Renyi entropy~\cite{Leone2022}. Although the latter quantifies the nonstabilizerness of a multi-qubit state, the expression in Eq.~(\ref{eqn:single_site_renyi}) directly looks at nonCliffordness of the transformation. As such, one might interpret it as the operator space counterpart to the stabilizer Renyi entropy, and we expect both quantities to have similar behaviors, that is, if for a multi-qubit state $V|\psi\rangle^{\otimes N}$ the stabilizer Renyi entropy is high/low, then Eq.~(\ref{eqn:single_site_renyi}) for some input Pauli string $O$ and the same unitary $V$ will be high/low.

The $P_l\in\mathcal{P}_N$ can be ordered by their ``weight'', i.e., the number of nonidentity elements in the Pauli string. This ordering allows us to systematically study the Clifford character of the transformation $V$ on Pauli strings. Naturally, the first step will be to check it for strings of weight one, which is done by setting $O=Y_n$, where $Y_n$ denotes a Pauli operator with a Pauli-$y$ on the $n$-th qubit position and identity everywhere else. In particular we denote $\mathcal{S}\left(Y_n;V\right) = \mathcal{S}_n\left(V\right)$. Since we can place the initial Pauli-$y$ at any of the $N$ positions representing the nodes of the graph, we consider the node-averaged Renyi entropy of the operator distribution
\begin{equation}
\label{eqn:mean_renyi}
\overline{\mathcal{S}}\left(V\right) = \frac{1}{N}\sum_{n=1}^N \mathcal{S}_n(V),
\end{equation}
as our figure of merit. Since Clifford unitaries map Pauli strings to Pauli strings,  then $\mathcal{S}\left(O;V\right) = 0$ for all $O\in\mathcal{P}_N$. Since we are only checking the behavior of $V$ as a ``channel'' for Pauli strings localized on one qubit, a vanishing $\overline{\mathcal{S}}$ is necessary (but not sufficient) for $V$ to be Clifford. We thus use  $\overline{S}$ as a witness of Cliffordness. 

We supplement this witness with an examination of the optimal parameters $(\boldsymbol{\gamma}^*, \boldsymbol{\beta}^*)$. Observation of $\boldsymbol{\gamma}^*, \boldsymbol{\beta}^* = \pm m \frac{\pi}{4}$ with $m\in \mathbb{N}$, then provides the sufficient condition for $V$ to be Clifford. This observation is made quantitative via the distance of the vector of parameters $\mathbf{v}$ to the discrete set of interest. We define this distance as 
\begin{equation}
\label{eqn:parameter_distance}
\mathcal{D}(\mathbf{v}) = \sum_{v_i\in \mathbf{v}} \frac{\underset{l\in\mathbb{Z}}{\rm min}[|v_i - l\frac{\pi}{4}|]}{\pi/8},
\end{equation}
where the normalization ensures that each term in the sum is bounded to the interval $[0,1]$, thus we have $0\le\mathcal{D(\mathbf{v})}\le|\mathbf{v}|$. Then the instance averaged distances $\mathbb{E}[\mathcal{D}(\gamma^*)]\to0$ and $\mathbb{E}[\mathcal{D}(\beta^*)]\to0$ will inform us of solution circuits which are close to Clifford.

\section{Origin of the ADAPT-Clifford algorithm}
\label{sec:algo_origin}
\begin{figure}[t!]
\centering{\includegraphics[width=0.99\linewidth]{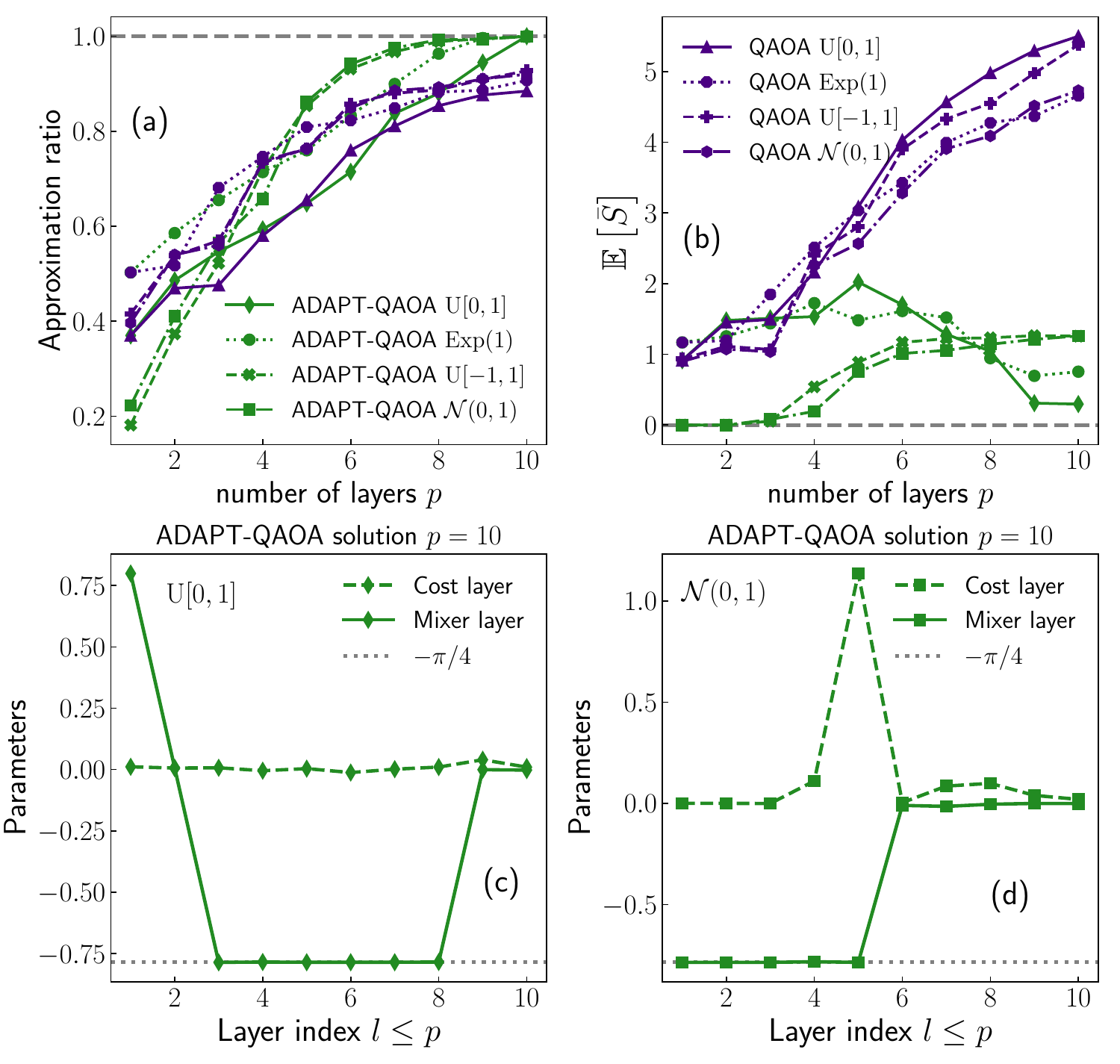}}
\caption{QAOA and ADAPT-QAOA results for MaxCut on weighted complete graphs with $N=6$. \textbf{(a)} Instance average of $\alpha$ for QAOA (purple) and ADAPT-QAOA (green). \textbf{(b)} Instance average of $\overline{\mathcal{S}}$ in Eq.~\ref{eqn:mean_renyi} of solution circuits with $p$ layers fore QAOA (purple) and ADAPT-QAOA (green). For both (a) and (b) the distributions of the weights used are indicated in the figure. \textbf{(c,d)} Examples of the parameters $\boldsymbol{\gamma}^*$ (cost, solid line) and $\boldsymbol{\beta}^*$ (mixer, dashed line) of the solution unitary $U(\boldsymbol{\gamma}^*,\boldsymbol{\beta}^*)$ for an instance with weights drawn from ${\rm U}[0,1]$ (c) and $\mathcal{N}(0,1)$ (d).}
\label{fig:figure_1}
\end{figure}

To understand the origin of the ADAPT-Clifford algorithm, it is instructive to examine the solution circuits obtained with QAOA and ADAPT-QAOA for MaxCut on small weighted complete graphs. We implemented both variational algorithms in the extensible Julia framework Yao.jl~\cite{Luo2020yaojlextensible} and use the COBYLA optimizer. The analysis of the operator distribution in the Pauil basis was implemented using QuantumOptics.jl~\cite{kramer2018quantumoptics}. 

We consider first the case of graphs with positive weights with $\omega_{i,j}$ from either ${\rm U}[0,1]$, where by ${\rm U}[a,b]$ we denote the uniform distribution in the interval $[a,b]$, or ${\rm Exp}(1)$, the exponential distribution with mean $1$. In Fig.~\ref{fig:figure_1}a we show the mean approximation ratios for $50$ problem instances with $N=6$ for circuits up to $p=10$ layers. Similar to the observation in Ref.~\cite{Zhu2022}, ADAPT-QAOA (green diamonds and circles in Fig.~\ref{fig:figure_1}a) finds a solution arbitrarily close to the exact solution at sufficiently high but finite $p$, away from the $p\to\infty$ limit where QAOA is guaranteed to reach the exact solution. In Fig.~\ref{fig:figure_1}b we show the expectation value over instances of $\overline{\mathcal{S}}$, $\mathbb{E}[\overline{\mathcal{S}}]$. The ADAPT-QAOA solution circuits that lead to $\alpha\to1$ in Fig.~\ref{fig:figure_1}a display $\mathbb{E}[\overline{\mathcal{S}}]\to0$, indicating they might be Clifford circuits. In contrast, the QAOA solution unitaries show $\mathbb{E}[\overline{\mathcal{S}}]>0$ with a tendency towards the typical value, $\log(4^{-N})$, with increasing depth, in agreement with previous works using other indicators~\cite{Valle2021,Dupont2022,Qian2023}. 

To verify the Cliffordness of the ADAPT-QAOA solution circuits we examine the optimized parameters, $(\boldsymbol{\gamma}^*,\boldsymbol{\beta}^*)$, at $p=10$. An example is shown in Fig.~\ref{fig:figure_1}c where dashed and solid lines correspond to $\boldsymbol{\gamma}^*$ and $\boldsymbol{\beta}^*$, respectively. We observe $\boldsymbol{\gamma}^* \to 0$ and $\boldsymbol{\beta}^*\to -\pi/4$ in all layers, and the mixer Hamiltonians selected by the adaptive step are almost always $Y_lZ_m$ for some pair of qubits $(l,m)$. Furthermore the distances of the optimal parameters for $p=10$ to $\pm s\frac{\pi}{4}$ with $s\in\mathbb{N}$ averaged over all instances with $\alpha\to1$ ($\omega_{i,j}\in {\rm U}[0,1]$), are $\mathbb{E}[\mathcal{D}(\gamma^*)] = 0.522 \pm 0.270$ and $\mathbb{E}[\mathcal{D}(\beta^*)] = 0.247 \pm 0.322$, indicating the optimized parameters are closed, on average, to the Clifford values. This is to be contrasted with $\mathbb{E}[\mathcal{D}(\gamma^*)] = 3.53 \pm 0.94$ and $\mathbb{E}[\mathcal{D}(\beta^*)] = 3.41 \pm 0.62$ for the optimized parameters of the QAOA solution circuits with $p=10$. Finally, we extensively checked that the properties of the ADAPT-QAOA solution unitary discussed here do not change as long as the edge weights are all positive.

Next we consider the case of signed weights with $\omega_{i,j}$ sampled either from ${\rm U}[-1,1]$ or $\mathcal{N}(0,1)$, the normal distribution with mean $0$ and variance $1$. In Fig.~\ref{fig:figure_1}a we compare the averaged approximation ratio of the ADAPT-QAOA solutions with that of the QAOA solutions for the same problem instances. Similar to the case of strictly positive weights, ADAPT-QAOA solutions get arbitrarily close to the exact solution when enough layers are considered. As seen in Fig.~\ref{fig:figure_1}b  $\mathbb{E}[\overline{\mathcal{S}}]\ne0$ for the ADAPT-QAOA solution (green crosses and squares). Although the circuits found are therefore not Clifford, $\mathbb{E}[\overline{\mathcal{S}}]\sim1$ at $p=10$ for the small problem size under study, in contrast to QAOA solution circuits (purple crosses and hexagons), for which $\mathbb{E}[\overline{\mathcal{S}}]$ tends towards the typical value. 

The small value of $\mathbb{E}[\overline{\mathcal{S}}]$ for the ADAPT-QAOA solutions raises the question: how far is this solution from the Clifford manifold? To answer this, in Fig.~\ref{fig:figure_1}d we show the optimized parameters $(\boldsymbol{\gamma}^*,\boldsymbol{\beta}^*)$ found for one of the problem instances solved. The $\boldsymbol{\beta}^*$\textquotesingle s are either $0$ or $-\pi/4$, indicating the mixer unitaries are Clifford, with mixer Hamiltonians almost always $Y_lZ_m$ for some pair of qubits $(l,m)$, and  most of the $\boldsymbol{\gamma}^*$\textquotesingle s are zero with only few, $\sim2$, being nonzero. Furthermore, the distances of the optimized parameters for $p=10$ to $\pm s\frac{\pi}{4}$ with $s\in\mathbb{N}$ averaged over all instances with $\alpha\to1$ ($\omega_{i,j}\in \mathcal{N}(0,1)$), are $\mathbb{E}[\mathcal{D}(\boldsymbol{\gamma}^*)] = 1.51 \pm 0.72$ and $\mathbb{E}[\mathcal{D}(\boldsymbol{\beta}^*)] = 0.83 \pm 0.96$, indicating that, on average, the solution circuits are further away from the Clifford manifold than in the case of $\omega_{i,j}>0$. This is to be contrasted with $\mathbb{E}[\mathcal{D}(\boldsymbol{\gamma}^*)] = 2.77 \pm 0.78$ and $\mathbb{E}[\mathcal{D}(\boldsymbol{\beta}^*)] = 2.75 \pm 0.64$ for the optimal parameters of the QAOA solution circuits with $p=10$. Therefore, the overall structure of the mixer unitaries of the ADAPT-QAOA $U(\boldsymbol{\gamma}^*,\boldsymbol{\beta}^*)$ found for positive $\omega_{i,j}$ is still there when $\omega_{i,j}$ are signed, complemented with a nontrivial non-Clifford action of a few of the cost layers. We have checked that this structure is common to all ADAPT-QAOA solutions reaching $\alpha\to1$.

We summarize the observations of this section:
\begin{itemize}
    \item The mixer part of all layers is Clifford with parameters either $0$ or $-\pi/4$. The mixer Hamiltonian at a given step is of the form $Y_lZ_m$ for some pair of qubits $(l,m)$.
    \item The cost part of most layers acts trivially with parameters equal to $0$. 
    \item Only $N$ steps are required to find an approximated solution. Consequently, only $N$ mixer layers of the form described in the first point are needed.
\end{itemize}

\section{ADAPT-Clifford approximation algorithm for MaxCut}
\label{sec:clifford_algo}
A bit string $\mathbf{z}^*$ is a good approximate solution to MaxCut if $\alpha(\mathbf{z}^*)$ is as close to $1$ as possible. Thus, finding good approximate solutions to this problem using only Clifford circuits means to prepare a stabilizer state $|\Psi\rangle$ whose energy expectation satisfies $|\langle\Psi|H_{\rm C}|\Psi\rangle - E_{\rm min}^{\rm C}|\le\epsilon$, with $\epsilon$ a small positive constant ideally equal to $0$. A measurement in the computational basis then returns $\mathbf{z}^*$ with the desired value of $\alpha$. 

Consider the bit string $\mathbf{z}'$ which maximizes the cost in Eq.~(\ref{eqn:classical_cost}). A stabilizer state satisfying the conditions discussed above is 
\begin{equation}
\label{eqn:stab_example}
|\Psi'\rangle = \frac{1}{\sqrt{2}}\left(\left|\mbf{z}'\right\rangle - \left|\overline{\mathbf{z}'}\right\rangle\right), 
\end{equation}
where $\overline{\mathbf{z}'}$ is the complement of $\mathbf{z}'$, and we have chosen the state to be antisymetric under the Ising symmetry $[H_{\rm C}, X^{\otimes N}]=0$, of the cost Hamiltonian. The state $|\Psi'\rangle$ is completely determined by its $N$ stabilizers. One of them is $-X_1 X_2 X_3...X_N$, while the remaining $N-1$ ones are of $ZZ$ type and their signs encode the maximal cut of the graph. In this setting, an approximation algorithm based on Clifford circuits must be able to determine an assignment of the signs of the $ZZ$ stabilizers leading to either $\mathbf{z}'$ or a $\mathbf{z}^*$ with $\alpha(\mathbf{z}^*)$ as close to one as possible.

\subsection{Details of the algorithm}
We design ADAPT-Clifford so as to exploit the observations summarized at the end of Sec.~\ref{sec:algo_origin} to prepare a stabilizer state $|\Psi\rangle$ with the general form given in Eq.~(\ref{eqn:stab_example}). This is done in a greedy manner, where after a choice of an initial seed, at every step the best local update is performed. As such, at an intermediate step $0<r \le N$ we label qubits as active and inactive. A qubit is active if a Pauli gate has been applied to it, otherwise it is inactive, and $a^{(r)}\in\mathbf{a}^{(r)}$ and $b^{(r)}\in\mathbf{b}^{(r)}$ are indices denoting the positions of ``active'' and ``inactive'' qubits, respectively, and $\mathbf{a}^{(r)}$ and $\mathbf{b}^{(r)}$ are vectors storing the positions of all the active and inactive qubits at step $r$.

ADAPT-Clifford prepares $|\Psi\rangle$ starting from the $k$-th qubit and growing this entangled state qubit by qubit, in such a way that at step $r$ the state is a product of two parts: an entangled state of all the $|\mathbf{a}^{(r)}|$ active qubits and all the $|\mathbf{b}^{(r)}|$ inactive qubits in the product state $|+\rangle^{\otimes |\mathbf{b}^{(r)}|}$. To specify the pair $(a^{(r)},b^{(r)})$ of qubit indices at each step, we use a ``gradient'' criterion similar to that of ADAPT-QAOA. Specifically, at step $r>2$ we compute
\begin{multline}
\label{eqn:gradients}
g_{a^{(r-1)},b^{(r-1)}}^{(r)} = -i\langle[H_{\rm C},Z_{a^{(r-1)}} Y_{b^{(r-1)}}]\rangle_{r-1} \\
 = -\sum_l \omega_{l,b^{(r-1)}}\langle Z_l X_{b^{(r-1)}} Z_{a^{(r-1)}} \rangle_{r-1},
\end{multline}
where $\langle.\rangle_{r-1} = \langle\psi_{r-1}|.|\psi_{r-1}\rangle$ is taken on the state at step $r-1$. Then, we choose the pair of qubits $(a^{(r-1)},b^{(r-1)})$ that maximizes $g_{a^{(r-1)},b^{(r-1)}}^{(r)}$. The case of $r=1$ is special, and we discuss it below alongside the steps of the algorithm.

ADAPT-Clifford returns a candidate maximal cut $\mathbf{z}^*$ of a graph $\mathcal{G}$ after completing the following $N$ steps:

\begin{enumerate}[start=0]
    \item At step $r=0$ we begin by selecting a position $k$ and preparing the product state
    \begin{equation}
        \label{eqn:state_step_zero}
        |\psi_0\rangle = Z_k \mathrm{H}^{\otimes N}|0\rangle^{\otimes N}.
    \end{equation}
    At this point the active and inactive qubits are $\mathbf{a}^{(0)}=\{k\}$ and $\mathbf{b}^{(0)}=\{1,..,N\}\backslash \{k\}$.
    
    \item At step $r=1$, given that $a^{(0)}=k$ we can estimate the largest gradient analytically. In fact, $\underset{b^{(0)}}{\rm max}[g^{(1)}_{k,b^{(0)}}] = \underset{b^{(0)}}{\rm max}[\omega_{k,b^{(0)}}]$, thus the pair we are looking for is the edge $(k,j)$ of $\mathcal{G}$ with 
    \begin{equation}
        j = \underset{b^{(0)}}{{\rm argmax}}[\omega_{k,b^{(0)}}].
    \end{equation}
    After applying the gate $e^{i{\frac{\pi}{4}}Y_k Z_j}$, the state is 
    \begin{equation}
        |\psi_{1}\rangle = e^{i\frac{\pi}{4}Z_j Y_k} Z_k \mathrm{H}^{\otimes N}|0\rangle^{\otimes N}.
    \end{equation}
    The vectors of active and inactive qubits are updated to $\mathbf{a}^{(1)}=\{k,j\}$ and $\mathbf{b}^{(1)}=\{1,..,N\}\backslash \{k,j\}$, respectively.
    
    \item For $r=2,...,N-1$, we find the pair of qubits $(\tilde{l},b^{(r-1)})$, with $\tilde{l}\in\{k,j\}$, which maximizes $g^{(r)}_{\tilde{l},b^{(r-1)}}$, apply the gate $e^{i{\frac{\pi}{4}} Z_{\tilde{l}} Y_{b^{(r-1)}}}$, and update the vectors of active and inactive qubits. In the case of more than one pair $(\tilde{l},b^{(r-1)})$ leading to the same largest value of $g^{(r)}_{a^{(r-1)},b^{(r-1)}}$ we break the tie arbitrarily. 

    \item After all $N$ steps are completed, we perform a measurement in the computational basis. From the output bit string, $\mathbf{z}_{\rm out}$, we readout the approximate maximal cut of the graph as $(\mathcal{A},\overline{\mathcal{A}})$ with $\mathcal{A} = \{ z_i \in \mathbf{z}_{\rm out} | z_i = 0, i=1,..,N\}$ and $\overline{\mathcal{A}} = \{ z_i \in \mathbf{z}_{\rm out} | z_i = 1, i=1,..,N\}$. 

    
\end{enumerate}
After the above $N$ steps are completed, the resulting stabilizer state $|\Psi\rangle$ encoding the solution has the form
\begin{equation}
\label{eqn:adapt_clifford_state}
|\Psi\rangle = \left[ \prod_{r=2}^{N-1} e^{i\frac{\pi}{4} Z_{\tilde{l}} Y_{b^{(r)}}} \right] e^{i\frac{\pi}{4} Z_{j} Y_{k}} Z_k \mathrm{H}^{\otimes N}|0\rangle^{\otimes N}.
\end{equation}

While it may seem that restricting the search to pairs of the form $(\tilde{l},b^{(r-1)})$ in step 2 may lead to missing the true largest gradient, in App.~\ref{app:proof_statement} we show that this is not the case. Furthermore, this restriction has a simple interpretation. After step $r=1$, we have effectively selected the edge $(k,j)$ as a reference with respect to which we are going to partition the graph. Nodes $k$ and $j$ are thus representatives of the disjoint subsets of the cut. Thus, from that step onward, we can pick $a^{(r-1)}\in\{k,j\}$ without loss of generality in order to decide which qubit to move into the active set, i.e., to include in the entangled state.

Some further comments are in order: (i) Given the type of two-qubit gate we are considering, the form of the initial product state $|\psi_0\rangle$ is chosen as to guarantee that $\underset{b^{(0)}}{\rm max}[g^{(1)}_{k,b^{(0)}}]$ will be positive. (ii) For $r>1$, and independently of the graph connectivity, not all the terms in the sum in Eq.~(\ref{eqn:gradients}) are nonzero; in fact, the expectation values in $g^{(r)}_{a^{(r-1)},b^{(r-1)}}$ become $\langle Z_l X_{b^{(r-1)}} Z_{a^{(r-1)}} \rangle_{r-1} = \langle Z_l Z_{a^{(r-1)}} \rangle_{r-1}$ and are nonzero only for those values of $l$ for which either $\pm Z_lZ_{a^{(r-1)}}$ is a stabilizer of $|\psi_{r-1}\rangle$. This observation allow us to find the largest gradient without explicitly computing the expectation values in Eq.~\ref{eqn:gradients}, which we show in App.~\ref{app:simple_gradient}. At the same time, this observation establishes a direct connection between ADAPT-Clifford and a family of existing MaxCut euristics~\cite{Sahni1976,Kahruman2007}, as was recently pointed out in Ref.~\cite{Wang2023}. (iii) The relevant two-qubit gate can be written in terms of Clifford gates as 
\begin{equation}
\label{eqn:clifford_form_gates}
e^{i{\frac{\pi}{4}}Y_l Z_m} = \mathrm{S}_l \mathrm{H}_m {\rm CNOT}_{\rm l,m} R_x^{(l)}(-\pi/2) {\rm CNOT}_{l,m}\mathrm{S}^\dagger_l \mathrm{H}_m, 
\end{equation}
where the $\mathrm{S}_l$, $\mathrm{H}_l$, are the phase and Hadamard gates acting on the $l$-th qubit, ${\rm CNOT}_{l,m}$ is the controlled NOT gate, with qubit $l$ and qubit $m$ as control and target qubits, respectively. Furthermore one can write $R_x^{(l)}(-\pi/2) = \mathrm{H}_l^{YZ} Z_l$ with $\mathrm{H}_l^{YZ}$ a variant of the Hadamard gate which swaps the $y$- and $z$-axes. For the interested reader, we work through the operations of our algorithm for two small examples in App.~\ref{app:examples}.

\subsubsection{A stabilizer perspective on the algorithm}
We can gain further understanding of the inner workings of the algorithm by looking at the way the stabilizers of the state change from step $r=0$ to step $r=N-1$. At step $r=0$, the product state $|\psi_0\rangle$ has $N-1$ stabilizers equal to $X_{l}$, $l=1,..,N$, $l\ne k$ and the remaining stabilizer equal to $-X_{k}$. At step $r=1$ the action of the gate between qubits $(k,j)$, with $j$ found as described previously, increases the weight of the $-X$ stabilizer by one and changes one of the $+X$ stabilizers by a $ZZ$ stabilizer. The state $|\psi_1\rangle$ is hence stabilized by $-\mathbb{I}_1..X_k\mathbb{I}_{k+1}...\mathbb{I}_{j-1}X_j..\mathbb{I}_N$ and $-\mathbb{I}_1..Z_k\mathbb{I}_{k+1}...\mathbb{I}_{j-1}Z_j..\mathbb{I}_N$ while the remaining $N-2$ are still $X_l$ with $l\ne k,j$. This process continues until $r=N-1$; with every new gate the weight of the $-X$ stabilizer increases by one and one of the $+X$ stabilizers gets replaced by a $ZZ$ stabilizer. We see then that the Clifford gate $e^{i\frac{\pi}{4}Y_lZ_m}$ was not chosen arbitrarily. In fact ADAPT-QAOA finds it because it is the gate that maps $X_l$ to $Z_lZ_m$.

As such we can phrase goal of the algorithm to be the correct assignment of the signs of the $ZZ$ stabilizers. After all $N-1$ steps are completed, the state $|\psi_{N-1}\rangle$ has one stabilizer equal to $-X_1 X_2 X_3...X_N$ and the remaining $N-1$ stabilizers are $ZZ$ with signs that were determined in the previous steps. If this sign assignment is done correctly, it encodes the approximate maximal cut produced by the algorithm. One can read it out directly by setting the value of any spin to either $+1$ or $-1$ arbitrarily and use the measured signs of the $ZZ$ stabilizers to fix the values of the spins at the other $N-1$ positions relative to the first one.
 
\subsection{Runtime and space complexities}
Although finding the qubit with the largest gradient at a given step does not require the explicit computation of expectation values of Pauli strings (see App.~\ref{app:simple_gradient} for details), it is defined based on a double sum with indices running on portions of the vertex set. As such, this part of the algorithm incurs the leading runtime cost.

At step $r>1$ and before applying the two-qubit gate, there are $r-1$ active qubits and $N-r+1$ inactive qubits. In order to decide on which pair of qubits we act the gate, we compute Eq.~(\ref{eqn:gradients}) via Eq.~(\ref{eqn:simplified_gradient}) for all pairs $(\tilde{l},b^{(r-1)})$ where $\tilde{l}\in\{k,j\}$ and $b^{(r-1)}\in\mathbf{b}^{(r-1)}$. There are $2(N-r+1)$ of those pairs. For a given pair the sum in Eq.~(\ref{eqn:gradients}) is $\forall_l$ such that $(l,\tilde{l})\in \mathcal{E}$. However only when $l\in\mathbf{a}^{(r-1)}$ is the expectation value $\langle Z_l X_b^{(r-1)} Z_{\tilde{l}}\rangle_{r-1}$ nonzero. Hence, at step $r$, there are $\delta = {\min}(r-1, K)$ with $K$ the maximum degree of the graph, nonzero terms in the sum. For bounded-degree graphs, such as $K$-regular graphs, $\delta=O(K)$ at most, whereas for dense graphs with $K=O(N)$, $\delta=O(N)$.

For a fixed initial position $k$, the algorithm executes $N-1$ steps before reaching a candidate solution. The total number of nonzero terms involved in the computation of the largest gradients is $\sum_{r=2}^{N-1}2(N-r+1)\delta$, so we have
\begin{equation}
2K\sum_{r=2}^{N-1}(N-r+1) = K(N^2-N+2), \nonumber
\end{equation}
or
\begin{equation}
2\sum_{r=2}^{N-1}(N-r+1)(r-1) = \frac{2}{3}(N^3-7N+6), \nonumber 
\end{equation}
for bounded-degree and dense graphs, respectively. Leading to a run time complexity of $O(N^2)$ for bounded degree graphs and $O(N^3)$ for dense graphs.

Since in general the initial position $k$ leading to the best approximate solution is not known, we propose and explore two complementary approaches. In the first approach, we choose the initial position at random. This algorithm, to which we refer as randomized ADAPT-Clifford, leads to run time complexities of $O(N^2)$ and $O(N^3)$ for bounded-degree and dense graphs, respectively, as described above. Second, we introduce a deterministic version ---deterministic ADAPT-Clifford--- where the best initial position $k^*$ is determined by exhaustive search. That is, we run ADAPT-Clifford $N$ times, each with a different initial position $k$, and return the cut of minimal energy found. The runtime complexity of this deterministic approach is thus $O(N^3)$ and $O(N^4)$ for bounded-degree and dense graphs, respectively. Naturally, the deterministic approach is guaranteed to return solutions of equal or smaller energy expectation that the randomized approach, at the cost of a more limiting runtime. Whether there exist graph families for which any initial position is as good as any other is a question for future work. Finally, it is easy to see that for both randomized and deterministic approaches the space complexity of the algorithm is $O(N^2)$, corresponding to the memory required to store the Tableau.

\section{Algorithm performance on weighted complete graphs}
\label{sec:performance_on_complete}
We have implemented the ADAPT-Clifford algorithm using the fast stabilizer circuit simulator Stim~\cite{Gidney2021stimfaststabilizer}. Our implementation is available at~\cite{ADAPT_Cliff_implementation}. Although we have chosen this simulator to implement our algorithm, any stabilizer circuit simulator which supports interactivity, that is, where expectation values of Pauli strings can be computed and the circuit modified according to the results, could be used to implement the algorithm.

We follow the presentation of Sec.~\ref{sec:algo_origin} and discuss separately our algorithm\textquotesingle s performance for MaxCut on weighted complete graphs with positive and signed weights. For the latter case, we will focus on the Sherrington-Kirkpatrick model.

\subsection{The case of positive weights}
\label{subsec:positive_weight}
\begin{figure}[t!]
\centering{\includegraphics[width=0.99\linewidth]{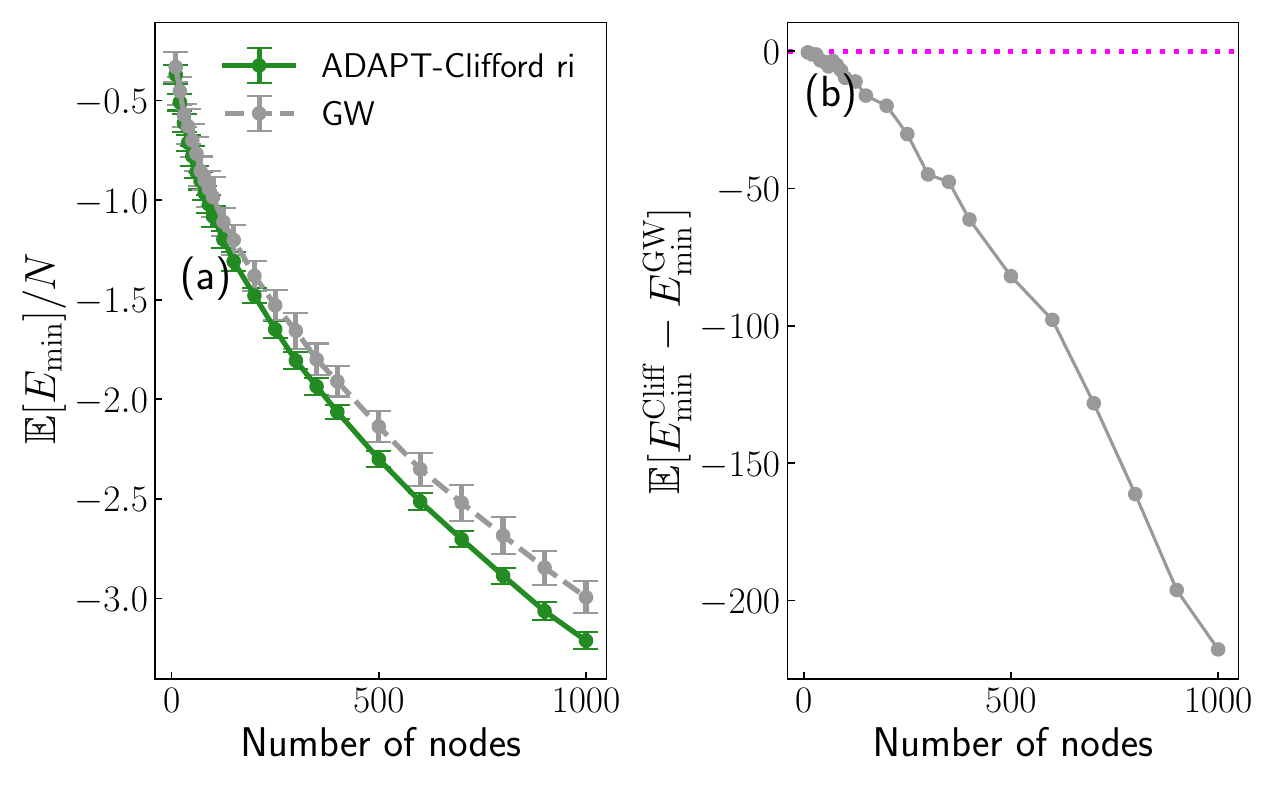}}
\caption{Performance of randomized ADAPT-Clifford on weighted complete graphs. \textbf{(a)} Normalized energy found by randomized ADAPT-Clifford (green circles) and GW (light grey circles) averaged over $100$ instances. \textbf{(b)} Instance averaged minimum energy difference between the solution found with randomized ADAPT-Clifford and GW as a function of problem size. Notice that randomized ADAPT-Clifford is almost always superior to GW. The magenta dotted line indicates a mean energy difference of zero. We have omitted the error bars for the sake of clarity.}
\label{fig:figure_2_1}
\end{figure}

The results of Sec.~\ref{sec:algo_origin} indicate that the precise choice of positive weight distribution may be immaterial. We have verified numerically that this is indeed the case for a few different weight distributions. In this subsection, we focus the discussion to $\omega_{i,j}$ sampled from ${\rm U}[0,1]$ and leave an exhaustive investigation for future work.

We begin studying the performance of the randomized approach. We draw a parallel between the random initialization of ADAPT-Clifford and the rounding step of GW, and thus assess the performance of the randomized ADAPT-Clifford by direct comparison with GW. We solved $100$ different problem instances for graph sizes up to $N=1000$ with both algorithms. In Fig.~\ref{fig:figure_2_1}a we show the normalized mean minimum energy, $\mathbb{E}[E_{\rm min}]/N$, of the solutions obtained with randomized ADAPT-Clifford (green circles) and the ones obtained with GW (light grey circles). Notice that our randomized ADAPT-Clifford almost always produces a solution of lower energy expectation than GW. These observations can be further verified with the mean difference of the minimum energy found, $\mathbb{E}[E_{\rm min}^{\rm Cliff} - E_{\rm min}^{\rm GW}]$, which we show in Fig.~\ref{fig:figure_2_1}b. Since our randomized ADAPT-Clifford consistently beats GW, we expect it to have a performance guarantee for typical instances of positively weighted complete graphs above that of GW for the general problem. We discuss the methodology used to estimate it in App.~\ref{app:estimate_mean_ratio}. We find $\overline{\alpha}^{\rm r}\approx0.8986$ a value which confirms our intuition and sets a lower bound for the expected performance of the deterministic ADAPT-Clifford.

\begin{figure}[t!]
\centering{\includegraphics[width=0.99\linewidth]{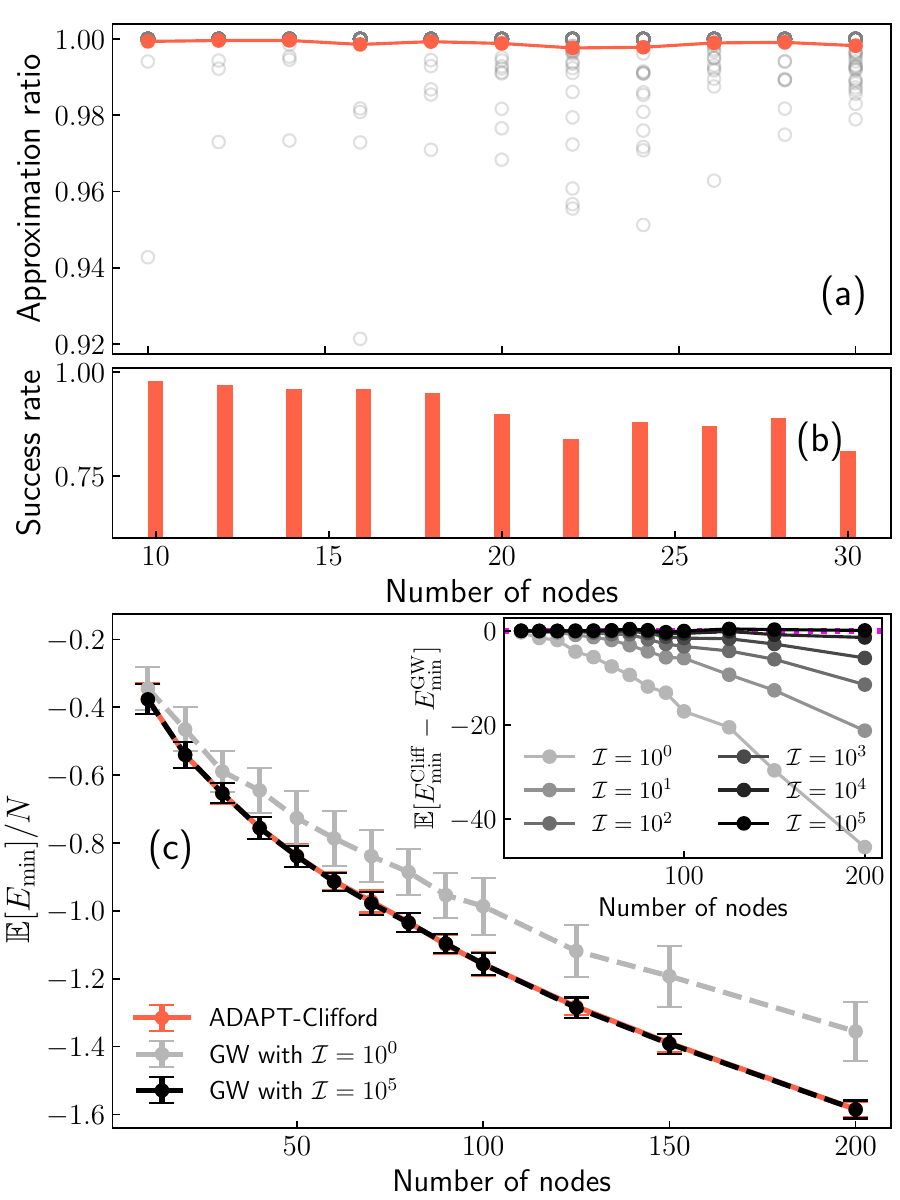}}
\caption{\textbf{(a)} Approximation rations $\alpha$ (empty circles) and mean approximation ratios (orange full circles) of solutions to MaxCut on $100$ weighted complete graphs per graph size found with deterministic ADAPT-Clifford. \textbf{(b)} Success rate on the $100$ problem instances per graph size considered in (a). \textbf{(c)} Instance-averaged minimum energy over $60$ problems up to graphs with $200$ nodes, both whit our algorithm (orange solid line), with Goemans-Williamson algorithm (light grey dashed line), and Goemans-Williamson with $\mathcal{I}=10^5$ (black dashed line). The inset shows the mean difference in the minimum energies found by our algorithm and the GW algorithm as a function of the problem size and for different values of $\mathcal{I}$. The magenta dotted line indicates a mean energy difference of zero. In the inset we have omitted the error bars for the sake of clarity.}
\label{fig:figure_2}
\end{figure}

We now focus on the deterministic approach. First, we benchmark this algorithm for instances with size up to $N=30$ for which the exact solution can be found exhaustively. Fig.~\ref{fig:figure_2}a shows the exact approximation ratios $\alpha$, obtained for $100$ problem instances. For these small problems, our algorithm performs, on average, above $\alpha = 0.997$, with the value of the minimum $\alpha$ increasing as $N\to30$. We notice that the number of instances for which our algorithm finds the exact ground state slightly decreases with the problem size. The success rate, defined as the number of times the algorithm finds a cut with energy $E_{\rm min}^{\rm Cliff}-E^{\rm C}_{\rm min}<10^{-10}$, is shown in Fig.~\ref{fig:figure_2}b as a function of the problem size. We observe a success rate $\sim80\%$ for $N=30$.

For problem sizes beyond $N=30$, when we cannot access the exact value of the ground state energy, we resort to a direct comparison with the GW algorithm. We find that the cuts obtained with deterministic ADAPT-Clifford are of superior quality to those found with the standard GW algorithm. To obtain a comparison, we thus systematically increase the number of times $\mathcal{I}$ the rounding step is performed in GW and return the best cut found ---see App.~\ref{app:gw} for details. The standard GW algorithm thus corresponds to $\mathcal{I}=1$. In Fig.~\ref{fig:figure_2}c we show the normalized mean energies $\mathbb{E}[E_{\rm min}]/N$ for $60$ problem instances up to a problem size of $N=200$ produced by our algorithm (orange circles), standard GW (light grey circles), and GW with $\mathcal{I}=10^5$ (black circles). Notice that our algorithm produces cuts which are \emph{always}, not merely on average, better than those produced with standard GW, and only when we reach $\mathcal{I}=10^5$ does the GW algorithm begin to produce a cut whose quality is, on average, superior to that of the cut produced by our algorithm. 

To further verify this observation, the inset of Fig.~\ref{fig:figure_2}c shows the mean energy difference, $\mathbb{E}[E_{\rm min}^{\rm Cliff} - E_{\rm min}^{\rm GW}]$, between the solution found with our algorithm and the one found with GW, with the magenta dotted line indicating $\mathbb{E}[E_{\rm min}^{\rm Cliff} - E_{\rm min}^{\rm GW}] = 0$, that is, equal quality cuts on average. It is seen that ADAPT-Clifford performs increasingly better than GW with fixed $\mathcal{I}$ as problem size is increased. To quantify the approximation quality of ADAPT-Clifford, we estimate the average approximation ratio of the deterministic ADAPT-Clifford on this family of graphs to be $\overline{\alpha}=0.9686$ -- see App.~\ref{app:estimate_mean_ratio} for details.    

To complement our benchmarks we performed a time to solution (TTS) study of GW with variable $\mathcal{I}$, randomized ADAPT-Clifford and deterministic ADAPT-Clifford, the details are shown in App.~\ref{app:TTS}. While Fig.~\ref{fig:figure_2}c shows that $\mathcal{I}=10^5$ rounding steps are needed for the GW algorithm to match the approximation quality of the deterministic ADAPT-Clifford for problem sizes up to $N=200$, the data in the inset imply that $\mathcal{I}$ may in fact need to scale with $N$ for the GW algorithm to compete with ADAPT-Clifford. However, in App.~\ref{app:TTS} we show that the TTS of GW remains constant with $I$ up to $I=10^3$ and only slightly increased for larger values of $\mathcal{I}$, and even at $\mathcal{I}=10^5$ the TTS of GW is faster than deterministic ADAPT-Clifford. In fact, we empirically verify the $O(N^{3})$ runtime scaling of GW (see App.~\ref{app:TTS} and~\cite{Goemans1995,Alizadeh1995} and references therein).

However, we also find an empiric runtime scaling for randomized ADAPT-Clifford of $O(N^{2.7})$, indicating an advantage. The comparison here is more involved, as GW with $\mathcal{I}>10^1$ already produces a better solution than randomized ADAPT-Clifford. However the Cholesky decomposition performed as part of GW is usually executed as a multi-core operation in most numerical linear algebra libraries. On the contrary, our implementation of ADAPT-Clifford runs on a single core, thus one could then easily improve the quality of solution without sacrificing the TTS or runtime scaling by executing the algorithm for different initial position in parallel. We thus believe, this variant of the algorithm does offer an advantage over GW.

\begin{figure}[t!]
\centering{\includegraphics[width=0.99\linewidth]{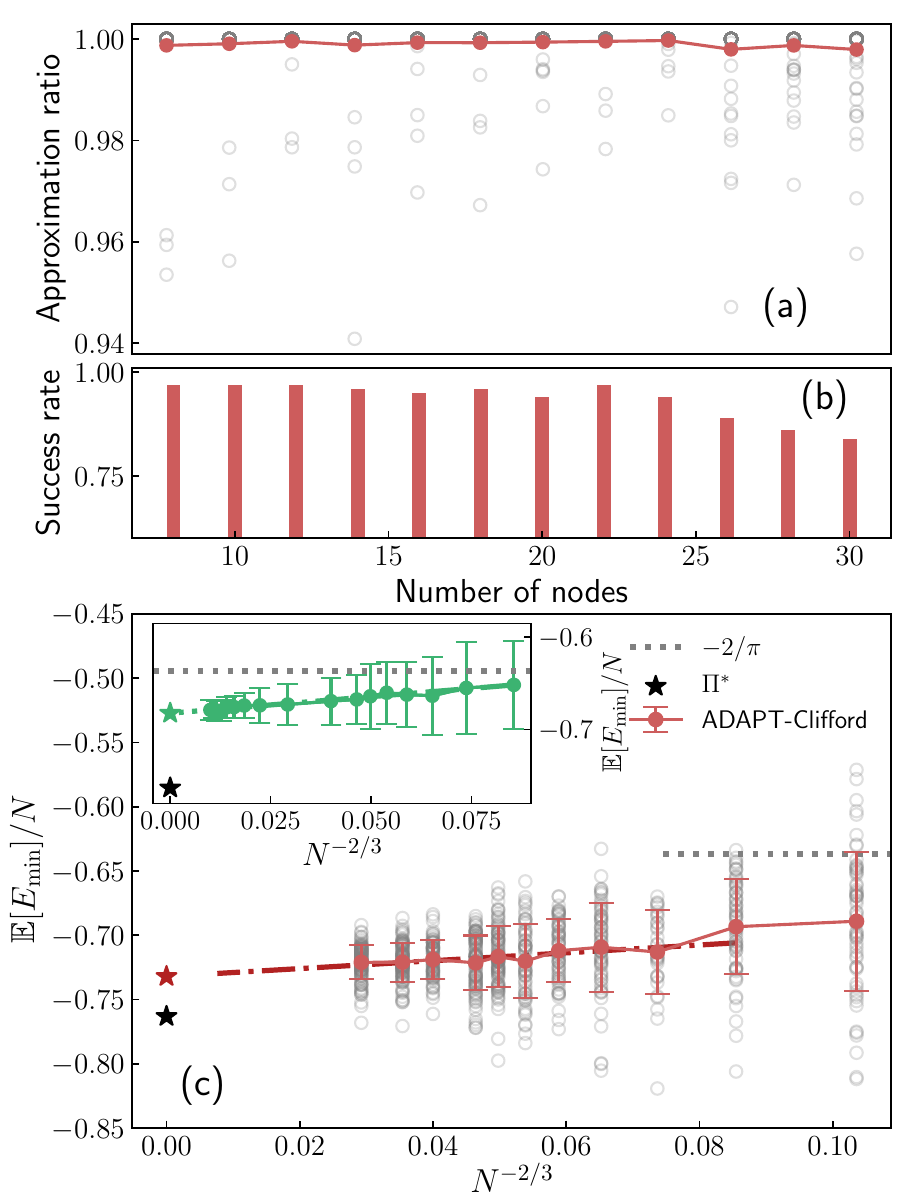}}
\caption{\textbf{(a)} Approximation rations $\alpha$ (empty circles) and mean approximation ratios (red full circles) of deterministic ADAPT-Clifford for $100$ different disorder realizations of the SK model per system size. \textbf{(b)} Success rate on the $100$ problem instances per graph size considered in (a). \textbf{(c)} Ground-state energy density for each of the $100$ problem instances (empty circles) per problem size up to $N=200$, and their mean (full circles). The dashed-dotted line show the best linear fit and the red star the respective mean energy density in the thermodynamic limit. The inset shows the average ground state energy density up to $N=1000$ as obtained with randomized ADAPT-Clifford, with its corresponding linear fit (see main text) and the value in the thermodynamic limit (green star). The grey dotted line shows the mean energy density obtained with semidefinite programing and the black star shows the Parisi value $\Pi^*$.}
\label{fig:figure_3}
\end{figure}

\subsection{Signed weights: the Sherrington-Kirkpatrick model}
\label{subsec:signed_weight}
The Sherrington-Kirkpatrick (SK) model~\cite{Sherrington1975} has played a fundamental role in the advancement of the understanding of the physics of spin glasses and disordered systems~\cite{Binder1986,Panchenko2012,Panchenko2013,Charbonneau2023}. It describes $N$ classical spins with all-to-all couplings of both ferromagnetic and antiferromagnetic character. The Hamiltonian is given by 
\begin{equation}
\label{eqn:sk_hamil}
H_{\rm sk} = \frac{1}{\sqrt{N}}\sum_{i<j}\omega_{i,j}\sigma_i\sigma_j,
\end{equation}
where $\sigma_i\in\{-1,1\}$ is a classical spin and the couplings $\omega_{i,j}$ are sampled from a distribution with zero mean and unit variance, for instance the normal distribution $\mathcal{N}(0,1)$. A milestone result by Parisi~\cite{Parisi1979,Mezard1986} gave an explicit expression for the ground state energy density of this model in the thermodynamic limit, which we refer to as the Parisi value, 
\begin{equation}
\label{eqn:parisi_value}
\lim_{N\to\infty}\mathbb{E}\left[\frac{E^{\rm sk}_{\rm min}}{N} \right] = \Pi^* = -0.763166...,
\end{equation}
where the expectation value is over realizations of the random couplings, and $E^{\rm sk}_{\rm min}$ refers to the ground state energy of Hamiltonian in Eq.~(\ref{eqn:sk_hamil}). The most accurate numerical value of Eq.~(\ref{eqn:parisi_value}) to date was computed in Ref.~\cite{Schmidt2008}. The limit in the LHS of Eq.~(\ref{eqn:parisi_value}) has been formally shown to both exist and be equal to the Parisi value~\cite{Guerra2003,Talagrand2006}.

Recently the SK model has been used as a benchmark in the study of quantum approximate optimization algorithms~\cite{Farhi2022,Dupont2023,Carlson2023}. Motivated by these works, we focus our attention on this model to characterize the performance of our algorithm on complete graphs with signed weights. A word of caution: The ADAPT-QAOA solution circuits for the signed case, including small instances of the SK model, are not completely Clifford, see Fig.~\ref{fig:figure_1}b,d and Sec.~\ref{sec:algo_origin}. As such, we do not expect our algorithm to match the solution quality of the best classical algorithm due to Montanari~\cite{Montanari2019,El2021}, which produces a $\boldsymbol{\sigma}^*$ with energy below $(1-\epsilon)$ times the lowest energy for typical instances, with $\epsilon$ a small positive constant~\footnote{The run time of this algorithm is $\mathrm{c}(\epsilon)O(N^2)$ with $\mathrm{c}(\epsilon)$ an inverse polynomial of $\epsilon$.}. Nevertheless, we are interested in seeing how close the $\boldsymbol{\sigma}^*$\textquotesingle s produced by our algorithm get to the Parisi value, both for the randomized and deterministic variants of ADAPT-Clifford.

\begin{figure*}[ht!]
\centering{\includegraphics[width=0.99\linewidth]{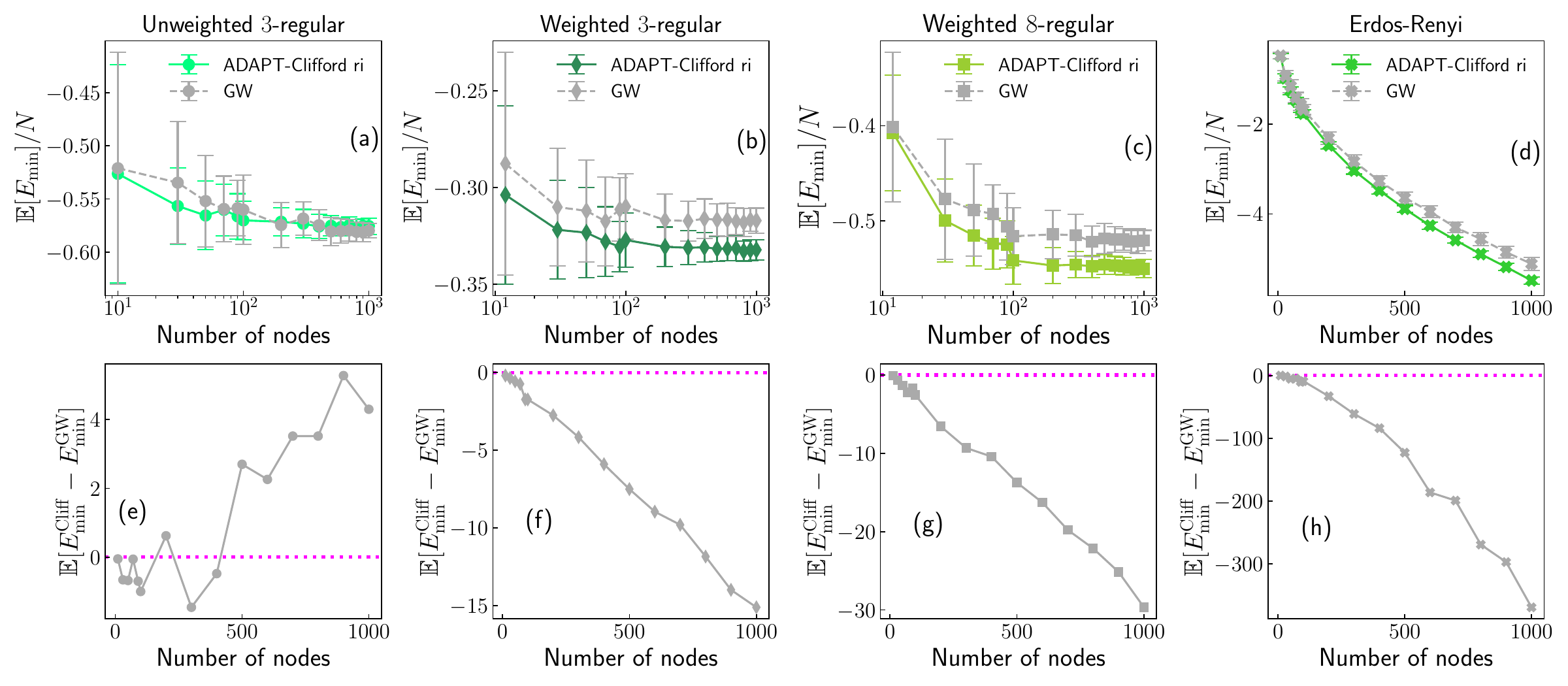}}
\caption{Performance of randomized ADAPT-Clifford vs GW. \textbf{(a-d)} Normalized instance-averaged minimum energy found with randomized ADAPT-Clifford (colorful markers and solid lines) and standard GW (light grey markers and dashed lines). The different graph types studied are: (a) unweighted $3$-regular graphs, (b) weighted $3$-regular graphs, (c) weighted $8$-regular graphs, (d) unweighted Erdös-Rényi graphs with edge probability $1/2$. For the weighted case we take $\omega_{i,j}$ in ${\rm U}[0,1]$. \textbf{(e-h)} Instance-averaged minimum energy differences between the solutions found with ADAPT-Clifford and standard GW for (e) unweighted $3$-regular (circles), (f) weighted $3$-regular (diamonds), (g) weighted $8$-regular (squares), and (h) Erdos-Renyi with edge probability $1/2$ (exes). The magenta dotted line indicates equal energy of the solutions found on average. We have omitted the error bars to avoid saturating the figure. All averages were computed over $100$ randomly generated instances.} 
\label{fig:figure_8}
\end{figure*}

In order to utilize ADAPT-Clifford we promote the classical spin in Eq.~(\ref{eqn:sk_hamil}) to $\sigma_i\to Z_i$ and use the resulting Hamiltonian as our cost. Following the presentation of the previous subsection, we discuss first the performance of the randomized ADAPT-Clifford. The green circles in the inset of Fig.~\ref{fig:figure_3}c show $\mathbb{E}\left[\frac{E_{\rm min}}{N} \right]$ for this algorithm with problem sizes up to $N=1000$. To obtain its value in the thermodynamic limit we fit the data for $N\in[40,1000]$ to a model of the form $qN^{-2/3} + \Pi^{\rm Cliff}_{\rm ri}$~\footnote{We point out that formal proof of the scaling $\mathbb{E}\left[\frac{E^{\rm sk}_{\rm min}}{N} \right]\sim N^{-2/3}$ only exists above the critical temperature, but it is believed that it holds for the whole spin glass phase. Thus assumption is supported with extensive numerical results, see for instance Ref.~\cite{Giberti2008,Palassini2008} and references therein. This is why we feel confident in its use in our model to fit the data.} where $\Pi^{\rm Cliff}_{\rm ri}$ corresponds to the mean energy density in the thermodynamic limit obtained with the randomized ADAPT-Clifford. We find $\Pi^{\rm Cliff}_{\rm ri} \approx -0.682$ which corresponds to $\sim89\%$ of the Parisi value (black star in inset of Fig.~\ref{fig:figure_3}c). This value is below what is obtained with convex relaxation methods, for instance semidefinite programming, which is known to give $\mathbb{E}\left[\frac{E_{\rm min}}{N} \right] = -\frac{2}{\pi} + o(1)\approx -0.6366$ with $o(1)$ a number which vanishes for $N\to\infty$~\cite{Aizenman1987,Montanari2016}. For comparison we display $\mathbb{E}\left[\frac{E_{\rm min}}{N} \right] = -\frac{2}{\pi}$ as the horizontal dotted line both in the inset and in Fig.~\ref{fig:figure_3}c.

Let us now consider the deterministic ADAPT-Clifford. For small problems $N\in[10,30]$ we computed the exact approximation ratios $\alpha$ over $100$ problem instances, and show them as empty circles in Fig.~\ref{fig:figure_3}a. Notably we do not observe $\alpha<0.94$ for any instance, and the average over instances is always above $\alpha>0.997$. To complement this observation we compute the success rate, defined as the number of instances for which the difference $E^{\rm SK}_{\rm min} - E_{\rm min}^{\rm Cliff} < 10^{-10}$. These are shown in Fig.~\ref{fig:figure_3}b with the smallest one being $\sim82\%$ at $N=30$.

To fully explore the performance of the deterministic ADAPT-Clifford algorithm, we solve $100$ instances for problems up to $N=200$. The normalized energies $E_{\rm min}^{\rm Cliff}/N$ are shown as empty circles in Fig.~\ref{fig:figure_3}c for all the instances considered, the red full circles show the respective $\mathbb{E}[E_{\rm min}^{\rm Cliff}]/N$ and the error bars correspond to the standard deviation of the normalized energies. To assess the quality of the solutions found we consider the data in the interval $N\in[40,200]$ and fit it to a model of the form $qN^{-2/3} + \Pi^{\rm Cliff}$ with $\Pi^{\rm Cliff}$ the estimated mean energy density in the thermodynamic limit of the solutions found by our algorithm. In particular for $N=200$ we find $\mathbb{E}\left[\frac{E^{\rm Cliff}_{\rm min}}{N} \right]\approx -0.727...$ and from the linear fit we find $\Pi^{\rm Cliff} \approx -0.7409...$, shown by a red star in Fig.~\ref{fig:figure_3}c. These values correspond to $\sim94\%$ and $\sim97\%$ of the Parisi value, respectively (the latter is shown with a black star in Fig.~\ref{fig:figure_3}c). These values are below what is obtained with convex relaxation methods (horizontal dotted line in Fig.~\ref{fig:figure_3}c). Notably, the value reached by our algorithm for $N=200$ is already better than what can be obtained with zero-temperature simulated annealing which gives $\mathbb{E}\left[\frac{E_{\rm min}}{N} \right]\sim-0.71$ (as quoted in Ref.~\cite{Farhi2022}), and $\Pi^{\rm Cliff}$ is comparable to what is achievable with simulated annealing on large problem instances.

\section{Algorithm performance on other families of graphs}
\label{sec:limitations}
\begin{figure}[t!]
\centering{\includegraphics[width=0.99\linewidth]{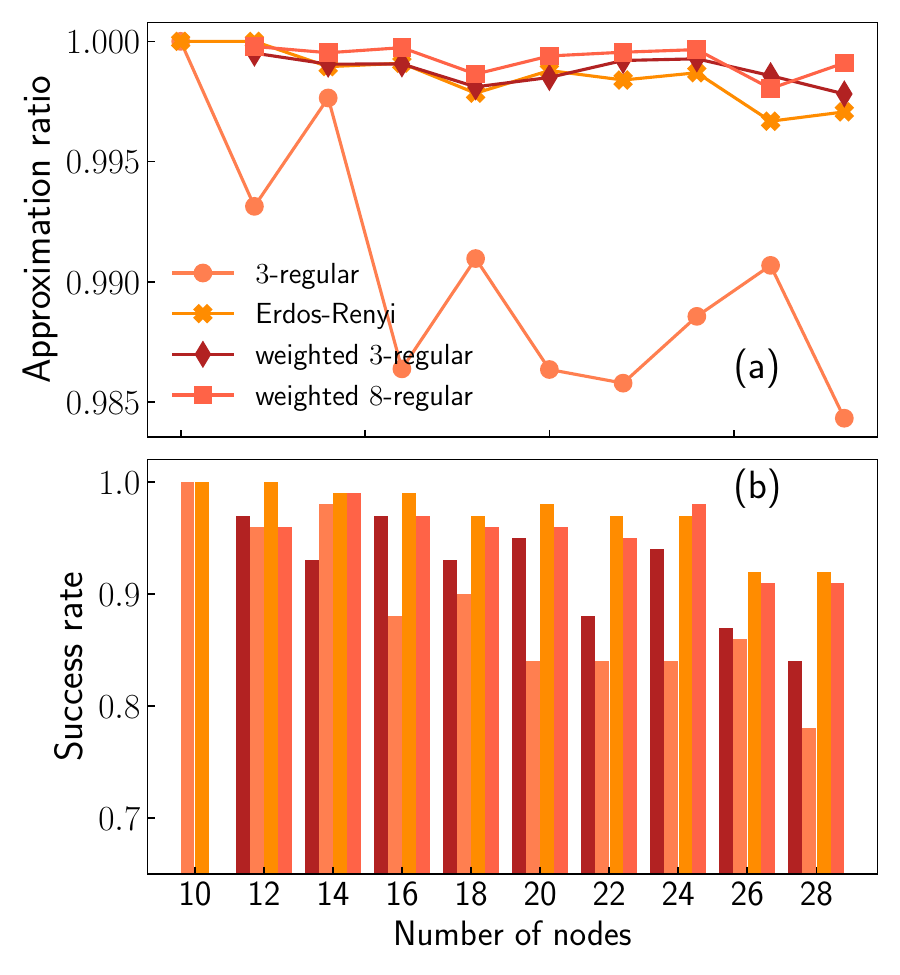}}
\caption{\textbf{(a)} Instance averaged approximation ratios up to $N=28$ for $100$ different instances of: unweighted $3$-regular graphs (circles), weighted $3$- (diamonds) and $8$-regular graphs (squares) with $\omega_{i,j}\in{\rm U}[0,1]$, and Erdös-Rényi graphs with edge probability $1/2$ (exes). We have omitted the error bars for clarity. \textbf{(b)} Success rate for each of the problem sizes and graph types considered in (a).}
\label{fig:figure_5}
\end{figure}
In this section, we characterize the performance of ADAPT-Clifford in both its variants for the MaxCut problem on $K$-regular graphs (unweighetd and weighted) and unweighted Erdos-Renyi graphs with various edge probabilities. For the randomized ADAPT-Clifford, we directly compare the quality of the cuts found with standard GW, while for determinisitic ADAPT-Clifford we discuss the exact approximation ratios for small problems and compare against GW with variable $\mathcal{I}$, the number of time the rounding step is performed.

\subsection{Performance on \texorpdfstring{$K$}{\textit{p}}-regular graphs}
\label{subsec:performance_on_regular}
We consider $3$-regular and $8$-regular graphs, unweighted and weighted. In all cases edge weights are sampled from ${\rm U}[0,1]$. 

In Fig.~\ref{fig:figure_8}a-c we show the normalized instance-averaged minimum energy of the solutions found with randomized ADAPT-Clifford and standard GW. For large ($N>200$) unweighted $3$-regular graphs, GW finds better solutions, on average, than randomized ADAPT-Clifford ---see Fig.~\ref{fig:figure_8}e. The situation is markedly reversed with the inclusion of nontrivial edge weights, with randomized ADAPT-Clifford outperforming standard GW (see Fig.~\ref{fig:figure_8}b), and the performance margin widens with increased connectivity, see Fig.~\ref{fig:figure_8}c. These observations are verified with the averaged minimum energy differences shown in Fig.~\ref{fig:figure_8}f,g for weighted $3$- and $8$-regular graphs, respectively. Thus, GW performs better than randomized ADAPT-Clifford only for unweighted $3$-regular graphs, while the comparative performance of our algorithm consistently improves with both the inclusion of edge weights and higher connectivity.

We now move to the performance of deterministic ADAPT-Clifford. In Fig.~\ref{fig:figure_5}a we show the mean approximation ratios over $100$ problem instances for each of these types of graphs. For the unweighted $3$-regular graphs we consider problem sizes $N\in[10,28]$ and for the weighted problems we consider problem sizes $N\in[12,28]$. We have omitted the error bars from the figure for the sake of clarity. Deterministic ADAPT-Clifford shows the poorest performance for unweighted $3$-regular graphs, circles in Fig.~\ref{fig:figure_5}a, with a decreasing mean $\alpha$ as $N$ increases. Interestingly the comparative performance of ADAPT-Clifford improves upon inclusion of edge weights, diamonds in Fig.~\ref{fig:figure_5}a, with a mean $\alpha$ above $0.995$ for all problem sizes considered. Further improved performance is observed with higher edge connectivity, as evidence by the mean approximation ratio for weighted $8$-regular graphs (squares in Fig.~\ref{fig:figure_5}a). In Fig.~\ref{fig:figure_5}b we show the success rate of the algorithm, i.e., the number of times ADAPT-Clifford found the maximal cut. The $3$-regular graphs (unweighted and weighted) show a success rate which consistently decay with problem size. On the contrary for weighted $8$-regular graphs our algorithm shows a success rate above $90\%$ up to $N=28$.
\begin{figure*}[t!]
\centering{\includegraphics[width=0.99\linewidth]{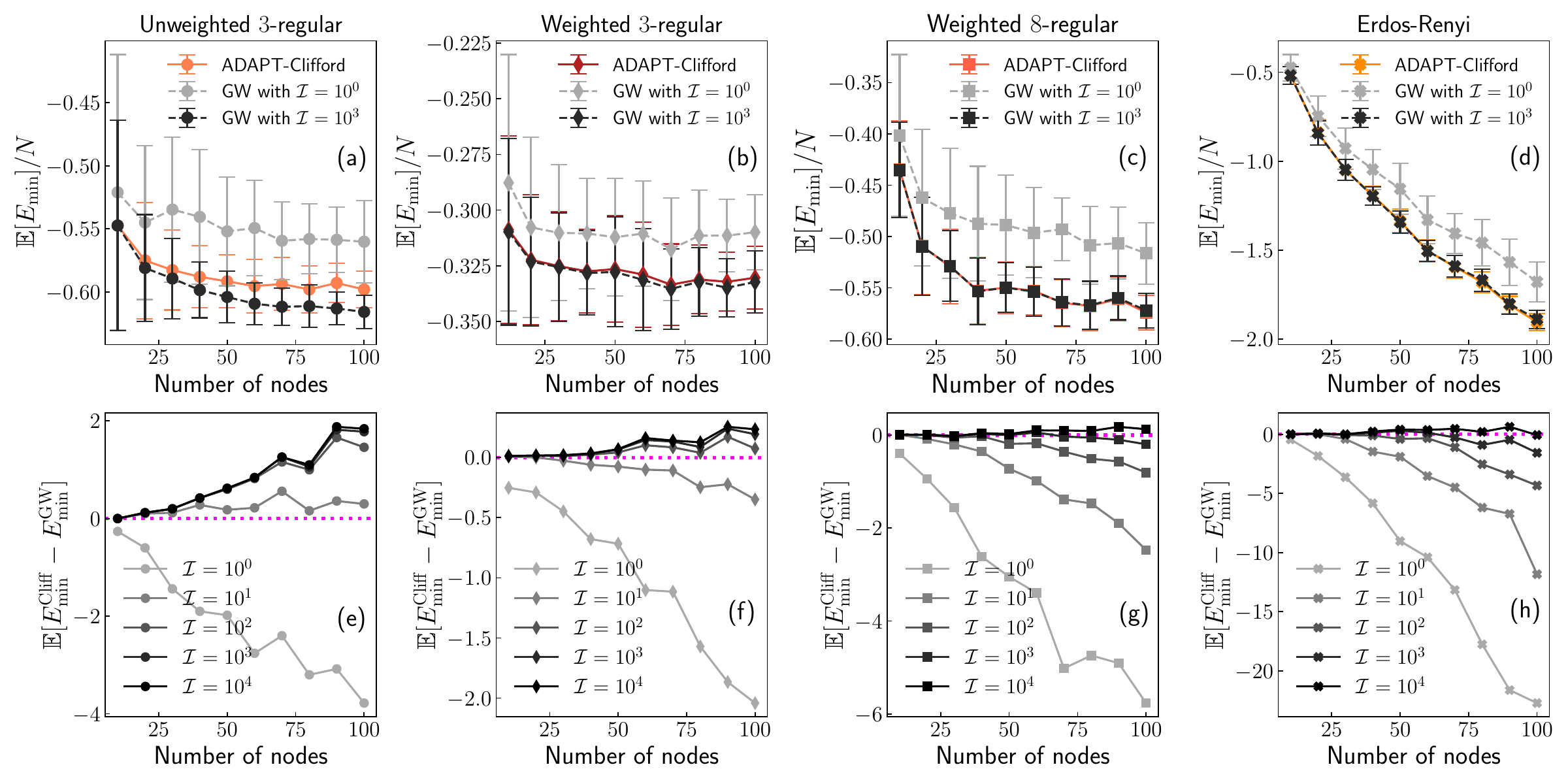}}
\caption{\textbf{(a-d)} Normalized instance-averaged minimum energy found over $100$ instances for problem sizes up to $N=100$ and different graph types obtained with deterministic ADAPT-Clifford (colorful markers and solid lines), with standard GW (light grey markers and dashed lines), and GW with $\mathcal{I}=10^3$ (dark grey markers and dashed lines). The different graph types studied are: (a) unweighted $3$-regular graphs, (b) weighted $3$-regular graphs, (c) weighted $8$-regular graphs, (d) unweighted Erdös-Rényi graphs with edge probability $1/2$. For the weighted case we always take $\omega_{i,j}$ in ${\rm U}[0,1]$. \textbf{(e-h)} Instance averaged minimum energy differences between the solutions found with our algorithm and the solution found with GW with different values of $\mathcal{I}$. For the graph types: (e) unweighted $3$-regular (circles), (f) weighted $3$-regular (diamonds), (g) weighted $8$-regular (squares), and (h) Erdos-Renyi with edge probability $1/2$ (exes). As a reference the the magenta dotted line indicates equal energy of the solutions found on average. We have omitted the error bars to avoid saturating the figure.} 
\label{fig:figure_4}
\end{figure*}

For larger problem sizes, we compare the solution quality of deterministic ADAPT-Clifford with that of GW with variable $\mathcal{I}$ ($\mathcal{I}=1$ is the standard GW algorithm). Fig.~\ref{fig:figure_4}a,b,c shows the normalized mean minimum energy $\mathbb{E}[E_{\rm min}]/N$ for the $K$-regular graphs studied. Notably, deterministic ADAPT-Clifford produces solutions of lower energy than GW for all three graph ensembles under consideration, compare the colorful markers with the light grey markers in Fig.~\ref{fig:figure_4}a,b,c. For the unweighted $3$-regular graphs, Fig.~\ref{fig:figure_4}a, already at $\mathcal{I}=10$ GW consistently finds a cut of lower energy than deterministic ADAPT-Clifford, signaling at a reduced performance of the latter method compared to the case of weighted complete graphs. This observation can be further verified with the mean difference of the minimum energy found, $\mathbb{E}[E_{\rm min}^{\rm Cliff} - E_{\rm min}^{\rm GW}]$, shown in Fig.~\ref{fig:figure_4}e. 

Similarly to its randomized counterpart, deterministic ADAPT-Clifford performs more competitively when edge weights are included. In Fig.~\ref{fig:figure_4}b we show the $\mathbb{E}[E_{\rm min}]/N$ obtained with our algorithm (red diamonds), GW (light grey diamonds), and GW with $\mathcal{I}=10^3$ (black diamonds), for the weighted $3$-regular graphs. Further inspection of the corresponding $\mathbb{E}[E_{\rm min}^{\rm Cliff} - E_{\rm min}^{\rm GW}]$, shown in Fig.~\ref{fig:figure_4}f, shows that at least $\mathcal{I}=10^2$ are necessary for the GW solution to be, on average, superior to that found by our algorithm. 
The performance margin widens as we move to regular graphs with higher connectivity. Fig.~\ref{fig:figure_4}c shows $\mathbb{E}[E_{\rm min}]/N$ obtained with deterministic ADAPT-Clifford (orange squares), standard GW (light grey squares), and GW with $\mathcal{I}=10^3$ (black squares), for weighted $8$-regular graphs. After inspecting the $\mathbb{E}[E_{\rm min}^{\rm Cliff} - E_{\rm min}^{\rm GW}]$ in Fig.~\ref{fig:figure_4}g we observe that $\mathcal{I}=10^4$ is necessary for the GW solution to be consistently better than the deterministic ADAPT-Clifford solution. Thus, for sparse graphs the performance of both randomized and deterministic ADAPT-Clifford improves with the inclusion of edge weights and/or higher node connectivity.

\subsection{Performance on unweighted Erdös-Rényi graphs}
\label{subsec:performamnce_er}

\begin{figure}[t!]
\centering{\includegraphics[width=0.99\linewidth]{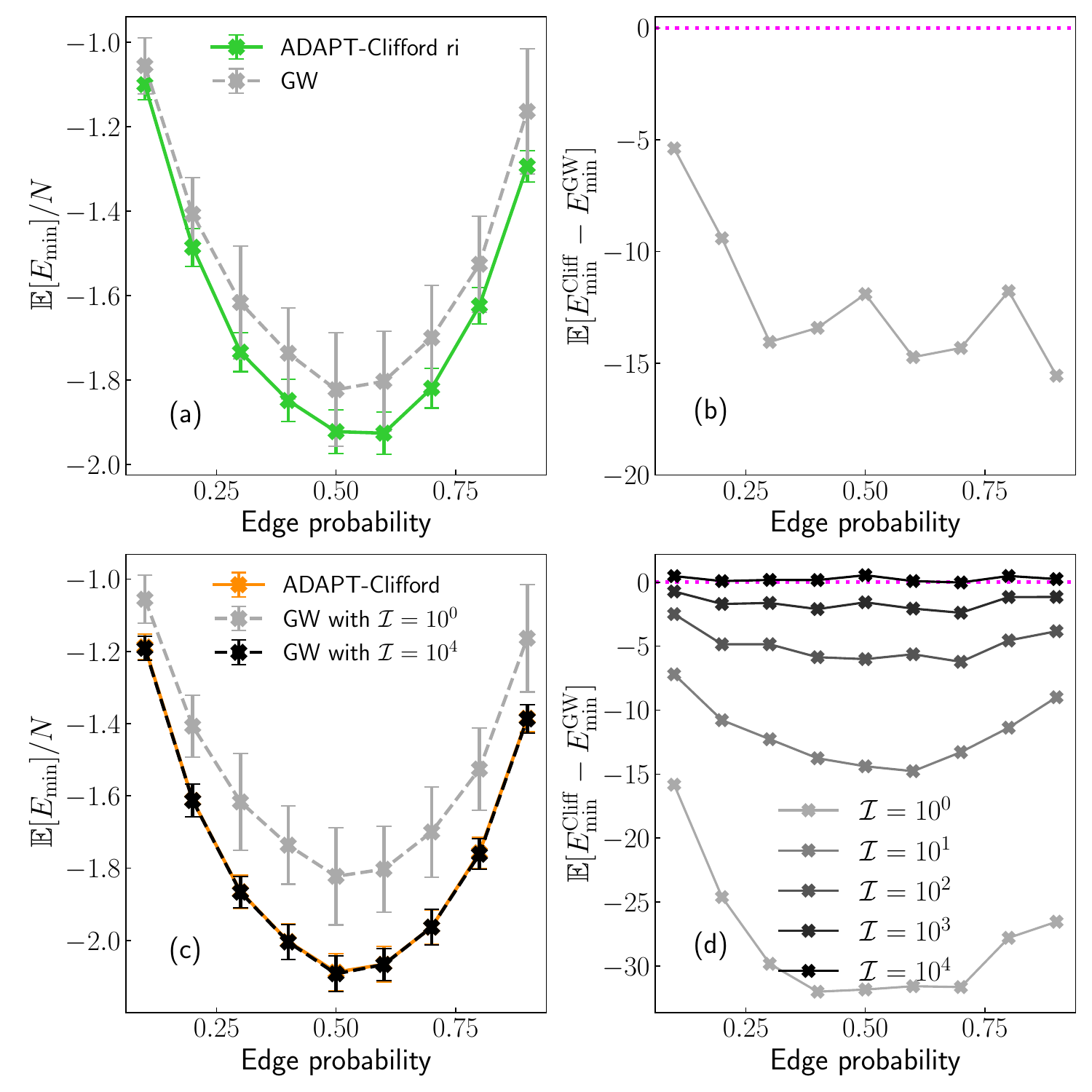}}
\caption{\textbf{(a)} Normalized instance-averaged minimum energy of the solutions found with randomized ADAPT-Clifford (green) and GW (light grey). \textbf{(b)} Instance averaged minimum energy differences between the solutions found with our algorithm and the solution found with GW. \textbf{(c)} Normalized instance averaged minimum energy of the solutions found with the deterministic ADAPT-Clifford (orange), standard GW (light grey), and GW with $\mathcal{I}=10^4$ (black), as a function of the edge probability which we take in $[0.1, 0.9]$. \textbf{(d)} Instance averaged minimum energy differences between the solutions found with our algorithm and the solution found with GW with different values of $\mathcal{I}$, as a function of edge probability. The magenta dotted line indicates mean energy difference of zero. The averages are taken over $100$ different problem instances and for $N=120$.}
\label{fig:figure_7}
\end{figure}

We now wish to characterize the performance of ADAPT-Clifford for MaxCut on dense graphs with variable density. For this task we will focus on unweighted Erdös-Rényi graphs.

First, we fixed the edge probability to $1/2$ and study the performance with respect to the problem size. In Fig.~\ref{fig:figure_8}d we show the instance averaged minimum energy of solutions obtained with randomized ADAPT-Clifford (green) and standard GW (light grey). For graphs up to $N=1000$. The randomized version of our algorithm produces better solutions, on average, than GW, an observation that is verified by the instance-averaged minimum energy differences shown in Fig.~\ref{fig:figure_8}h. 

Next, we analyze the performance of the deterministic ADAPT-Clifford on small problems $N\le28$. Fig.~\ref{fig:figure_5}a,b show mean approximation ratios (exes), which is above $\alpha\sim0.997$, and success rates, respectively. Notably, deterministic ADAPT-Clifford shows a higher success rate for this family of graphs, finding the maximal cut on all instances considered for the sizes $N=10,12$ (whereas it only achieves the same for the $19$ nonisomorphic $3$-regular graphs at $N=10$). For larger problems, in Fig.~\ref{fig:figure_4}d we compare the normalized instance-averaged minimum energy found by our algorithm (orange solid line), standard GW (light grey dashed line), and GW with $\mathcal{I}=10^3$ (black dashed line). Our algorithm (orange) finds a solution of lower energy, on average, than that found with GW (light grey). We explore the extent of this advantage by inspecting the mean difference of the minimum energy found, $\mathbb{E}[E_{\rm min}^{\rm Cliff} - E_{\rm min}^{\rm GW}]$, as a function of $N$ and with $\mathcal{I}$ as a control parameter. The results are shown in Fig.~\ref{fig:figure_4}h. It is seen that only at $\mathcal{I}=10^4$ the GW solutions are consistently of lower energy than those found by deterministic ADAPT-Clifford.

Now we turn our attention to benchmarking both the randomized and deterministic ADAPT-Clifford on Erdös-Rényi graphs with varying edge inclusion probability. We focus on problems with $N=120$ and consider edge probabilities in $[0.1, 0.9]$. We solve $100$ problem instances of MaxCut per edge inclusion probability. In Fig.~\ref{fig:figure_7}a we show the normalized mean energies found with the randomized ADAPT-Clifford (green) and with standard GW (light grey). Our randomized algorithm returns, on average, a cut of better quality than GW (see also instance-averaged minimum energy differences in Fig.~\ref{fig:figure_7}b). The normalized mean energies of the solutions found with deterministic ADAPT-Clifford (orange) are shown in Fig.~\ref{fig:figure_7}c, alongside those for standard GW (light grey), and GW with $\mathcal{I}=10^4$ (black). With the exception of edge probabilities smaller than $0.15$ and larger than $0.85$, the solutions found by our algorithm are \emph{always}, not merely on average, better than the ones found with GW, with the largest advantage observed for edge probabilities around $1/2$. Only at $\mathcal{I}\sim10^4$ does GW produce solutions on average comparable to those found by ADAPT-Clifford. This is seen more clearly in the instance-averaged energy difference of the solutions found $\mathbb{E}[E_{\rm min}^{\rm Cliff} - E_{\rm min}^{\rm GW}]$, shown in Fig.~\ref{fig:figure_7}b. Only at $\mathcal{I}=10^4$ we find $\mathbb{E}[E_{\rm min}^{\rm Cliff} - E_{\rm min}^{\rm GW}]\in(0, 0.5]$ for all edge probabilities, indicating our algorithm no longer offers an advantage over GW.

The results discussed here suggest that ADAPT-Clifford offers an advantage over GW on the quality of the cut found for dense graphs, with the largest gap for graphs with density $\sim1/2$. 

\section{Discussion and outlook}
\label{sec:outlook}
We introduce ADAPT-Clifford, a quantum inspired \emph{classical} approximation algorithm for MaxCut. For each problem instance, ADAPT-Clifford builds a low-depth Clifford circuit to prepare a stabilizer state that encodes an approximate solution. The algorithm was inspired by observation of the (almost) Clifford character of the ADAPT-QAOA solution circuits for MaxCut on weighted fully connected graphs. A comparison between the two methods and resource estimates for ADAPT-Clifford are given in App.~\ref{app:resource_estimate}. We introduce a randomized and a deterministic variant of this algorithm. Their respective runtime complexities are $O(N^2)$ and $O(N^3)$ for sparse graphs, and $O(N^3)$ and $O(N^4)$ for dense graphs, and in all cases the space complexity is $O(N^2)$. Naturally, the deterministic variant \emph{always} outperforms the randomized variant, albeit at the cost of an increased runtime. 

We have studied the performance of ADAPT-Clifford on MaxCut for various families of graphs, both dense and sparse, and both unweighted and weighted. On weighted complete graphs with positive weights, ADAPT-Clifford finds very high quality cuts, reaching the absolute maximum in the majority of small instances. Moreover, the algorithm is scalable, allowing us to easily find solutions for instances with up to $1000$ nodes. ADAPT-Clifford also performs well for signed weights, finding good approximations to the ground state of the SK model with an energy that extrapolates to $97\%$ of the Parisi value in the thermodynamic limit. To investigate performance as a function of density, we applied ADAPT-Clifford to MaxCut on unweighted Erdös-Rény graphs with variable edge inclusion probability. We again find that ADAPT-Clifford finds the absolute maximum cut for the majority of small instances and easily scales to hundreds of nodes. Finally, we study the performance of ADAPT-Clifford for sparse graphs. Even though these graphs are far from the context that gave rise to the algorithm, we find that ADAPT-Clifford still performs well, producing the absolute maximum cut with high probability for small instances and easily scaling to $1000$ nodes. Only for the case of $3$-regular graphs, the sparsest category of graphs we studied, we observe a noticeable deterioration in solution quality with increasing size. Counter-intuitively, performance improves somewhat with the inclusion of edge weights. 

To assess the performance of ADAPT-Clifford for large problem instances whose exact solution is intractable, we compare its performance with the GW algorithm, which represents the state of the art in approximate solution of MaxCut. For all graph families studied, ADAPT-Clifford outperforms the standard GW algorithm in the quality of the cut found. Only for very sparse unweighted graphs, such as $3$-regular graphs, the performance of the GW algorithm becomes comparable to that of ADAPT-Clifford, but even in this case the inclusion of edge weights favors the latter. Finally, ADAPT-Clifford solves problems to which the GW algorithm is not directly applicable, as exemplified by our results on the SK model.

The Clifford or near Clifford character of the ADAPT-QAOA solution circuits is a key observation which was missed in previous work~\cite{Zhu2022}. This observation, as laid out in Sec.~\ref{sec:algo_origin}, allowed us to devise a quantum-inspired, polynomial-time approximation algorithm for MaxCut. While it is known that MaxCut on dense graphs admits Polynomial Time Approximation Schemes (PTAS), leading to approximated solutions which are $1-\epsilon$ away from the optimum in time polynomial in $N$~\cite{Arora1999,de_la_vega2000}, the scaling of the runtime as a function of $\epsilon$ may render these algorithms impractical. In contrast, in this work we showed empirically that ADAPT-Clifford performs better than an algorithm that offers a guaranteed approximation ratio. Notably, based on the gradient criteria used as update rule in ADAPT-Clifford a connection between this algorithm and a family of heuristics for the MaxCut problem, known as Sahni-Gonzales algorithms~\cite{Sahni1976,Kahruman2007}, can be established, as was recently pointed out in Ref.~\cite{Wang2023}. It is remarkable to see that when ADAPT-QAOA performs best, the adaptive approach builds solution circuits which share this property with well known classical heuristics.

We hope the results reported here will help delimit the subset of graphs where a quantum speedup could be expected and thus where the current efforts should focus, in similar spirit to previous results obtained with a different subuniversal family of gates~\cite{Weitz2023sub}. While our work indicates that ADAPT-Clifford has a guaranteed approximation ratio, we do not yet have a proof. Our algorithm showed the poorest performance on fully connected graphs with signed weights. This was anticipated in Sec.~\ref{sec:algo_origin} since the ADAPT-QAOA solution circuits are not fully Clifford. However, they are \emph{near}-Clifford, motivating then a resource-centered design of variational ansätze, with a Clifford mixer part constructed following a scheme like the one introduced in this work, similar in spirit to the optimal mixers restricted to subspaces~\cite{Fuchs2023}, and a cost part with \emph{few} variational parameters adding just the right amount of nonCliffordness necessary to approximate the problem up to a desired ratio. Furthermore our algorithm could aid in reducing the cost of parameter optimization in QAOA when used to warm-start~\cite{Egger2021} it. More broadly, Clifford circuits can be leverage to construct a framework for the efficient state initialization in variational quantum algorithms beyond the product state paradigm. An example of this applied to quantum chemestry problems was introduced in Ref.~\cite{ravi2022cafqa}. Finally, our Clifford algorithm was tailored to solve the MaxCut problem. It remains an open question to what extent other combinatorial optimization problems admit Clifford approximation algorithms with practical polynomial runtimes.

\acknowledgments
The authors are grateful to Othmane Benhayoune-Khadraoui for helpful discussions, to Pablo Poggi for his comments in the early stages of the project, to Camille Le Calonnec for her insights on the implementation of adaptive variational quantum algorithms, to Maxime Dion for general discussion about the workings of the algorithm, and to Fen Zuo for pointing out to us the connection between ADAPT-Clifford and the Sahni-Gonzales family of MaxCut heuristics. This material is partially based upon work supported by the U.S. Department of Energy, Office of Science, National Quantum Information Science Research Centers, Quantum Systems Accelerator (QSA). Additional support is acknowledge from the Canada First Research Excellence Fund and the Ministère de l’Économie et de l’Innovation du Québec. SK is supported by a Research Chair in Quantum Computing by the Ministère de l'Économie, de l'Innovation et de l'Énergie du Québec.

\appendix
\section{The Goemans-Williamson algorithm}
\label{app:gw}
Suppose we are interested in solving the MaxCut problem for some given graph $\mathcal{G} = (\mathcal{V}, \mathcal{E})$ of $N$ nodes and edge weights $\omega_{i,j}$ using the Goemans-Williamson algorithm~\cite{Goemans1995,Goemans1995a}. To do so one proceeds as follows:

\begin{enumerate}
    \item Relax the binary character of the variables in the optimization problem defined by the cost function in Eq.~(\ref{eqn:classical_cost}), that is, replace the $z_i\in\{0,1\}$ with unit vectors $y_i\in\mathbb{R}^N$ and the product $z_iz_j$ with the inner product $y_i^Ty_j$ with $T$ the transpose. The new cost function $\sum_{i,j<i}\omega_{i,j}(1-y_i^Ty_j)$ with the constraints $y_i^Ty_i=1$, $\tilde{Y} = [y_i^Ty_j]$ is positive semidefinite, defines a semidefinite program.

    \item Solve the semidefinite program using a polynomial time algorithm, and find an optimal solution $\tilde{Y}^*$ for the relaxed problem.

    \item \textbf{Rounding}. Choose a random vector $\mathbf{r}\in\mathbb{R}^N$ from a Gaussian distribution and for all $i$ define $h_i = {\rm sgn}(\mathbf{r}^Ty^*_i)$, where ${\rm sgn}(x)$ is the sign function. This assignment defines a partition of the nodes in two disjoint sets $\mathcal{A} = \{i | h_i = 1\}$ and $\overline{\mathcal{A}} = \{i | h_i = -1\}$.

    \item return the cut $(\mathcal{A},\overline{\mathcal{A}})$.
\end{enumerate}
In this form the algorithm only performs the rounding, step (3), a single time based on a single random vector $\mathbf{r}$. As such, a simple improvement consists on repeating this step $\mathcal{I}$ times for different random vectors and then returning the cut of largest cost among all the cuts found. We have used this approach in comparing our algorithm with GW.

All the results for the GW algorithm reported in this manuscript have been obtained using a freely available Julia implementation~\cite{GW_implementation}.

\section{Validity of the search through a restricted set of pairs}
\label{app:proof_statement}
In step (2) of ADAPT-Clifford in Sec.~\ref{sec:clifford_algo}, we restricted our search to pairs of the form $(\tilde{l},b^{(r-1)})$ with $\tilde{l}\in\{k,j\}$ and $(k,j)$ is the edge where the first two-qubit gate was applied, and $b^{(r-1)}\in\mathbf{b}^{(r-1)}$. In this appendix we show than in doing so we do not miss the value of the largest gradient.

At step $r>1$ the gradient is of the form
\begin{multline}
\label{eqn:grad1_app}
g_{a^{(r-1)},b^{(r-1)}}^{(r)} = -\sum_l \omega_{l,b^{(r-1)}}\langle Z_l X_{b^{(r-1)}} Z_{a^{(r-1)}} \rangle_{r-1}\\
= -\sum_l \omega_{l,b^{(r-1)}}\langle Z_l Z_{a^{(r-1)}} \rangle_{r-1},
\end{multline}
where we have used the fact that $X_{b^{(r-1)}}|\psi_{r-1}\rangle = |\psi_{r-1}\rangle$ since $b^{(r-1)}$ is inactive. The maximum of Eq.~(\ref{eqn:grad1_app}) happens at the pair $(a^{(r-1)},b^{(r-1)})$ such that the number of $l$\textquotesingle s, with $l\in\mathbf{a}^{(r-1)}$, for which $-Z_l Z_{a^{(r-1)}}|\psi_{r-1}\rangle = |\psi_{r-1}\rangle$ is the largest. 

Now consider the situation of interest where we search for the pair to apply the gate among those of the form $(\tilde{l},b^{(r-1)})$, and suppose we know that the maximum of Eq.~(\ref{eqn:grad1_app}) occurs at the pair $(\tilde{a}, \tilde{b})$ with $\tilde{a}\in\mathbf{a}^{(r-1)}$ and $\tilde{b}\in\mathbf{b}^{(r-1)}$. Then
\begin{multline}
\label{eqn:grad2_app}
g_{\tilde{a},\tilde{b}}^{(r)} = \sum_l \omega_{l,\tilde{b}}\langle -Z_l Z_{\tilde{a}} \rangle_{r-1} \\
= \sum_l \omega_{l,\tilde{b}} \langle -Z_l Z_{\tilde{l}} Z_{\tilde{l}}Z_{\tilde{a}}\rangle_{r-1},
\end{multline}
where we introduced an identity $\mathbb{I}_{\tilde{l}} = Z_{\tilde{l}} Z_{\tilde{l}}$.

Since $\tilde{a}$ is active and $-Z_kZ_j|\psi_{r-1}\rangle = |\psi_{r-1}\rangle$ we can always pick the value of $\tilde{l}\in\{k,j\}$ such that $Z_{\tilde{a}}Z_{\tilde{l}}|\psi_{r-1}\rangle = |\psi_{r-1}\rangle$. Thus, we can extend Eq.~(\ref{eqn:grad2_app}) to the following chain of equalities
\begin{multline}
\label{eqn:grad3_app}
g_{\tilde{a},\tilde{b}}^{(r)} = \sum_l \omega_{l,\tilde{b}} \langle -Z_l Z_{\tilde{l}} Z_{\tilde{l}}Z_{\tilde{a}}\rangle_{r-1} \\ 
=\sum_l \omega_{l,\tilde{b}} \langle -Z_l Z_{\tilde{l}} \rangle_{r-1} = g_{\tilde{l},\tilde{b}}^{(r)}.
\end{multline}
We see then that the largest gradient does live within the restricted set of pairs of the form $(\tilde{l},b^{(r-1)})$.

\section{Gradient computation without explicit evaluation of the expectation values}
\label{app:simple_gradient}
As mentioned in step 2 of our algorithm description, from $r>1$ on-wards we select the qubit $b^{(r)}\in \mathbf{b}^{(r)}$ with the largest gradient with either $\{k,j\}$ and move it to the active set. The gradient is thus
\begin{equation}
g_{\tilde{l},b^{(r-1)}}^{(r)} = -\sum_l \omega_{l,b^{(r-1)}}\langle Z_l X_{b^{(r-1)}} Z_{\tilde{l}} \rangle_{r-1},
\end{equation}
where $\tilde{l}\in\{k,j\}$. Since $b^{(r-1)}$ is an inactive qubit $X_{b^{(r-1)}}|\psi_{r-1}\rangle = |\psi_{r-1}\rangle$. Further as was pointed out in Sec.~\ref{sec:clifford_algo} only those terms for which $l\in\mathbf{a}^{(r-1)}$ are nonzero. We write the gradient as  
\begin{equation}
g_{\tilde{l},b^{(r-1)}}^{(r)} = -\sum_{\substack{l\in\mathbf{a}^{(r-1)}\\
(l,b^{(r-1)})\in\mathcal{E}}} \omega_{l,b^{(r-1)}}\langle Z_l Z_{\tilde{l}} \rangle_{r-1}.
\end{equation}
Further note that the active qubits carry an additional label, indicating whether the qubit became active after being entangled with $k$ or with $j$. This label allow us to write the active qubits as $\mathbf{a}^{(r)}=V_{k}^{(r)} \cup V_j^{(r)}$ with $V_{k}^{(r)} \cap V_j^{(r)} = \emptyset$. Consider an inactive qubit at step $r-2$, importantly if it becomes active after being entangled with $k$, then the stabilizer generator at its position, $X_{b^{(r-2)}}$, changes to $e^{-i\frac{\pi}{4}Y_{b^{(r-2)}}Z_k} X_{b^{(r-2)}} e^{i\frac{\pi}{4}Y_{b^{(r-2)}}Z_k} = Z_{k}Z_{b^{(r-2)}}$. Since $-Z_kZ_j$ is a stabilizer of the state, then $-Z_jZ_{b^{(r-2)}}$ also stabilizes the state. Similarly, if the qubit becomes active after being entangled with $j$, then after application of the respective two qubit gate, the state is stabilized by both $Z_{j}Z_{b^{(r-2)}}$ and $-Z_{k}Z_{b^{(r-2)}}$. With these expressions we can immediately write the gradient as 
\begin{equation}
\label{eqn:simplified_gradient}
g_{k,b^{(r-1)}}^{(r)} = -\sum_{\substack{l\in\mathbf{V}_k^{(r-1)}\\
(l,b^{(r-1)})\in\mathcal{E}}} \omega_{l,b^{(r-1)}} + \sum_{\substack{l\in\mathbf{V}_j^{(r-1)}\\
(l,b^{(r-1)})\in\mathcal{E}}} \omega_{l,b^{(r-1)}},
\end{equation}
and $g_{j,b^{(r-1)}}^{(r)} = - g_{k,b^{(r-1)}}^{(r)}$. We thus see that the computation of the gradient does not require the explicit computation of the expectation values seen in Eq.~(\ref{eqn:gradients}).

We would like to point out that the explicit form of the gradient in Eq.~(\ref{eqn:simplified_gradient}) establishes a connection with a family of heuristic algorithms for MaxCut known by the name of Sahni-Gonzalez~\cite{Sahni1976,Kahruman2007}. This connection was recently studied in Ref.~\cite{Wang2023}.

\section{Time to solution analysis: ADAPT-Clifford vs Goemans-Williamson}
\label{app:TTS}
\begin{figure}[t!]
\centering{\includegraphics[width=0.99\linewidth]{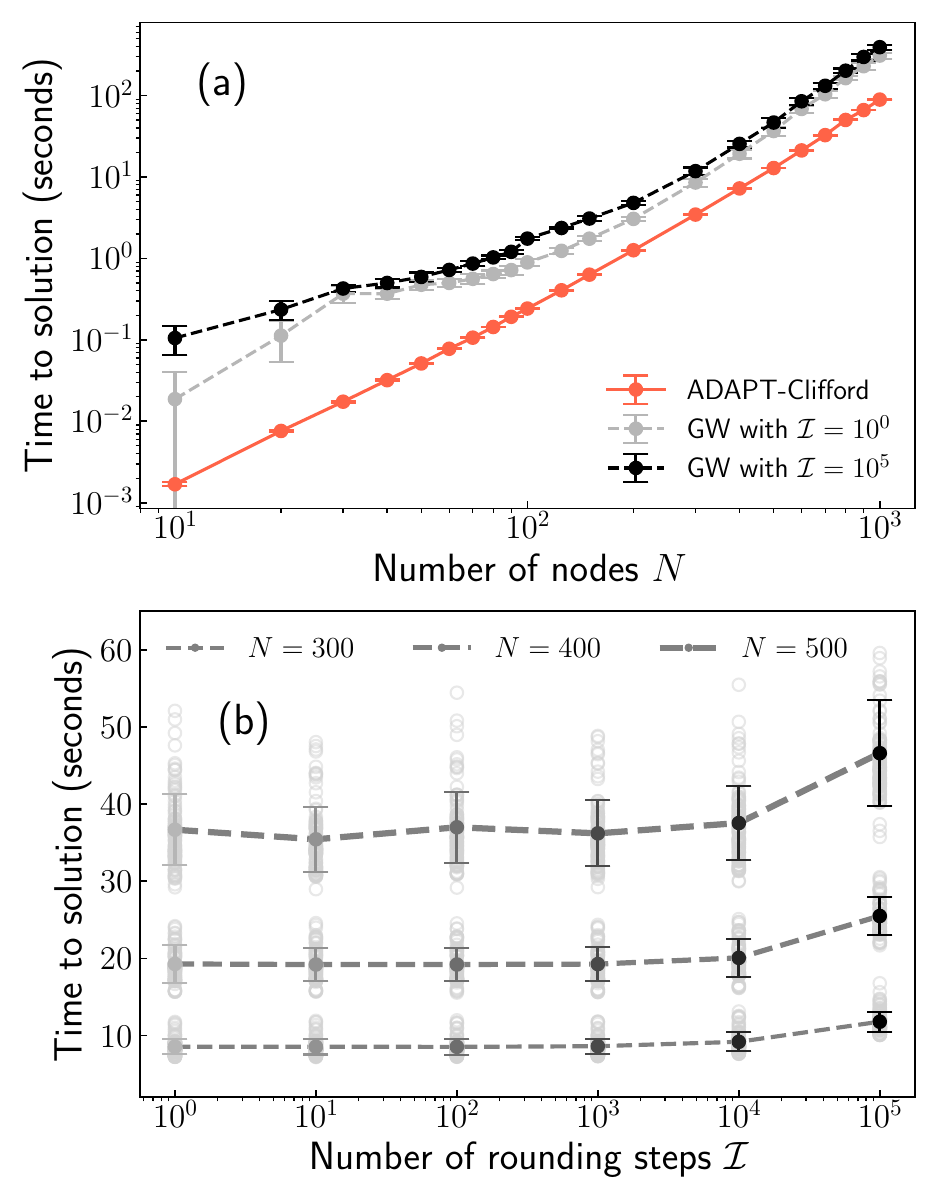}}
\caption{\textbf{(a)} Problem instance averaged time to solution for randomized ADAPT-Clifford (orange solid line) and GW with $\mathcal{I} = 10^0,10^5$ (dashed light grey and black), for the weighted complete graphs considered in Fig.~\ref{fig:figure_2_1}. \textbf{(b)} Time to solution of GW as a function of the number of times the rounding step $I$ is performed. Results are shown for three different system sizes, $N=300,400,500$ from bottom to top, respectively. The empty circles correspond to all the individual TTS for the different $100$ problem instances solved per problem size, the full circle to the mean TTS and the error bars shown the standard deviation. The dashed lines are guides for the eye.}
\label{fig:figure_9}
\end{figure}
In order to complete the comparison between the performance of ADAPT-Clifford and GW, we conducted a time to solution (TTS) analysis. This study was carried out on an Apple M1 Pro laptop with 8 cores. We focused on the case of weighted complete graphs studied in Sec.~\ref{subsec:positive_weight}, and compared our Python based implementation of ADAPT-Clifford available at~\cite{ADAPT_Cliff_implementation} against the freely available Julia implementation of GW for the MaxCut problem~\cite{GW_implementation}. Since the TTS of deterministic ADAPT-Clifford can be easily determined as one order of magnitude larger than that of randomized ADAPT-Clifford, we restricted the TTS study to the latter variant of our algorithm. For the GW we restricted ourselves to consider the situations of only $\mathcal{I}=10^0$ and $\mathcal{I}=10^5$, the two extreme values taken for the results in Sec.~\ref{subsec:positive_weight}.

The results are shown in Fig.~\ref{fig:figure_9}. Several observations are in order. The TTS of GW shows a relative slow increase with system size for small problems, until about $N\sim300$ after which the familiar $O(N^3)$ scaling is observed. Additionally, we do not find a strong dependence of the TTS with $\mathcal{I}$, in fact for the problem sizes considered in this work, \textit{i.e.}, up to $N=1000$, the TTS does not increase with $\mathcal{I}$ long as this is not bigger than $O(10^2)$, see Fig.~\ref{fig:figure_9}b. When the value of $I$ exceeds this threshold, we do observe an increase in the runtime albeit not a considerable one. As such, deterministic ADAPT-Clifford will only be superior both in solution quality and TTS up to $N=30$.

For randomized ADAPT-Clifford the TTS is always better than GW, see Fig.~\ref{fig:figure_9}a. In fact, we empirically find a more advantageous scaling of $O(N^{2.7})$ when contrasted to the $O(N^3)$ found in the same manner for GW. However, this makes the comparison considerably more subtle. Although the quality of solution found by GW is already better than randomized ADAPT-Clifford when $\mathcal{I}>10$, GW relies on a Cholesky decomposition which has highly optimized numerical subroutines that exploit the multi-core nature of modern CPUs. However, randomized ADAPT-Clifford is based on STIM and thus runs on a single core. One could then easily improve the quality of solution without sacrificing the TTS or runtime scaling by executing the algorithm for different initial position in parallel. We thus believe, this variant of the algorithm does offer an advantage over GW.

\begin{figure}[ht!]
\centering{\includegraphics[width=0.99\linewidth]{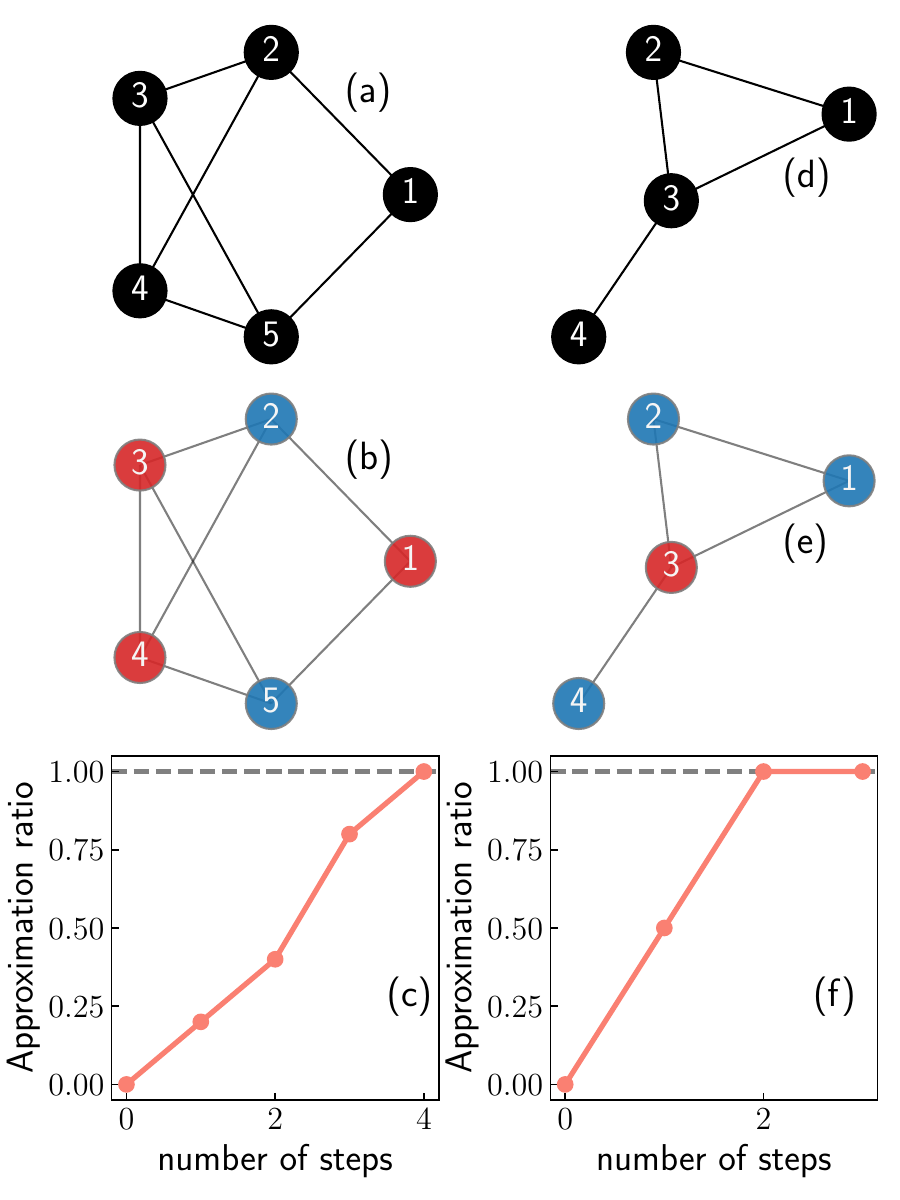}}
\caption{\textbf{(a,d)} Example graphs with $N=5$ and $N=4$ nodes, respectively. \textbf{(b,e)} Partitioned graphs according to the cuts produced by our algorithm, different colors denote the nodes in each of the disjoint subsets. \textbf{(c,f)} Approximation ratios of the states produced by our algorithm in the search process for the maximal cut of the graphs shown in (a) and (d). In both cases the algorithm reaches approximation ratio of $1$, indicating a maximal cut has been found.}
\label{fig:figure_examples}
\end{figure}

\section{Resource estimates for ADAPT-Clifford}
\label{app:resource_estimate}
As mentioned in Sec.~\ref{sec:outlook} one might wonder how ADAPT-Clifford compares with ADAPT-QAOA? And what would be the resource demands to implement ADAPT-Clifford on a near-term device? In this appendix we explore these two questions. 

Given the computational cost of directly simulating ADAPT-QAOA we restrict the comparison to small problem instances, as for example those solved in Fig.~\ref{fig:figure_1}. As we observed in Sec.~\ref{sec:algo_origin}, the best performing ADAPT-QAOA solution circuits are very close to Clifford. In this sense, the performance, quantified by the approximation ratio, of both ADAPT-QAOA and ADAPT-Clifford, for these small problem instances, must be the same. Now, if the comparison is extended other metrics, for instance number of layers or steps, cost of parameter optimization, cost of gradient evaluation, then ADAPT-Clifford will be best. 

In order to see this we will show that ADAPT-Clifford can be directly obtained from ADAPT-QAOA by systematically reducing the ansatz expressivity. We start from the full ADAPT-QAOA ansatz in Eq.~(\ref{eqn:adapt_state}), and use the observations summarized at the end of Sec.~\ref{sec:algo_origin} to systematically reduced the ansatz expressivity. First, we start from the state $Z_k\mathrm{H}^{\otimes N} |0\rangle^{\otimes N}$ where $k$ is chosen randomly. Second, we set all $\gamma_l=0$ and the $\beta_l=-\pi/4$, this guarantees the resulting circuit to be Clifford. Third, we restrict the operator pool in Eq.~(\ref{eqn:operator_pool}) to $\mathrm{P}_{\rm OP} = \{Y_l Z_m, Z_l Y_m\}_{j,k=1,...,N, j\ne k},$, where one of the two indices is fixed by our choice of $k$. Fourth, we use the gradient Criteria to select the mixer Hamiltonian at any intermediate step.

The above algorithm is a direct simplification of the full ADAPT-QAOA ansatz, and although it is different from the algorithm presented in Sec.~\ref{sec:clifford_algo}, it already allows us to see why ADAPT-Clifford is a more efficient algorithm. First, the gradient evaluation is highly simplified by the reduction of the operator pool size. Second, the need for parameter optimization is bypassed. Third, the restriction to Clifford unitaries enables efficient classical simulation. Finally, for ADAPT-Clifford we showed in App.~\ref{app:simple_gradient} that the gradient evaluation can be done without the need for explicit evaluation of expectation values, which further speeds up the algorithm.

Let us now focus on the second question. We based the resource estimation on the CNOT count. First, let us consider a qubit chip with all-to-all connectivity. In this case, and given the gate shown in Eq.~(\ref{eqn:clifford_form_gates}), it is easy to see that the total number of CNOT gates required to run the circuit is $2N$, with $N$ the number of nodes in the problem graph.

Consider now the opposite case, a qubit architecture with only linear connectivity. In order to apply a CNOT between two qubits at arbitrary positions, say $l$ and $m$, we bring the second qubit $m$ to $l+1$ using a swap network, then apply the Clifford gate, and swap the second qubit back to its original position $m$. In the following we use the fact that a swap can be implemented with $3$ CNOTs. Further, we assume that the two reference qubits, $k$ and $j$ are on opposite ends of the chain, thus $|k-j|=N-1$, and implementing the entangling Clifford gate between these two qubits requires $6(N-2) + 2$ CNOT gates. Finally, we assume the worst scenario in terms of separation between the qubits, where the entangling Clifford gates need to be applied between the reference qubit and the qubit which is furthest from it, among those inactive. Thus, the rest of the $N-2$ steps can be implemented using $2(N-2)+6\sum_{s=0}^{N-3}s = 3(N-3)(N-2) + 2(N-2)$ CNOT gates. Which leads to a total CNOT count of $2 + (N-2)\left[3(N-3) +8 \right]$.

\section{Some explicit examples of ADAPT-Clifford solving MaxCut}
\label{app:examples}
In this appendix, we go over the full analytical calculation of the steps involved in solving MaxCut using the algorithm introduced in Sec.~\ref{sec:clifford_algo} for two small graphs with $N=4,5$ nodes. In order to keep the expressions clean we have decided to focus on the case of unweighted graphs.

\subsection{An example with \texorpdfstring{$N=5$}{\textit{N=5}} nodes}
Consider the unweighted graph with five nodes shown in Fig.~\ref{fig:figure_examples}a. Its adjacency matrix is given by
\begin{equation}
[\omega_{i,j}] = \begin{pmatrix}
0 && 1 && 0 && 0 && 1 \\
1 && 0 && 1 && 1 && 0 \\
0 && 1 && 0 && 1 && 1 \\
0 && 1 && 1 && 0 && 1 \\
1 && 0 && 1 && 1 && 0
\end{pmatrix}.
\end{equation}
We will solve MaxCut on this graph using our algorithm. We begin by flipping the state of the qubit at $k=2$, thus we have 
\begin{equation}
|\psi_0\rangle = |+-+++\rangle,
\end{equation}
where $|+\rangle = \mathrm{H}|0\rangle$ and $|-\rangle = Z\mathrm{H}|0\rangle$ are the eigenstates of the Pauli-$x$ operator corresponding to eigenvalues $+1$ and $-1$, respectively. At this point we have to initialize the records of active and innactive qubits, which we identify by their respective indices. The active qubits are $[2]$ and the inactive qubits are $[1,3,4,5]$.

Given our choice of initial position, we have that $g_{1,2}^{(1)}=g_{2,3}^{(1)}=g_{2,4}^{(1)}=1$ and are the largest ``gradients''. We break this tie arbitrarily and chose the pair of qubits $(2,4)$. Then
\begin{equation}
|\psi_1\rangle = e^{i\frac{\pi}{4}Y_2 Z_4}|\psi_0\rangle = \frac{1}{\sqrt{2}}\left[ |+-+++\rangle - |+++-+\rangle \right],
\end{equation}
and the records of the active and inactive qubits are updated to be $[2,4]$ and $[1,3,5]$, respectively. The second set of gradients is given by 
\begin{widetext}
\begin{subequations}
\begin{align}
g_{1,2}^{(2)} &= -\sum_{l=2,5}\langle Z_lX_1Z_2 \rangle = -\langle X_1\rangle - \langle Z_5 X_1 Z_2 \rangle = -1 + 0 = -1, \\
g_{3,2}^{(2)} &= -\sum_{l=2,4,5} \langle Z_lX_3Z_2 \rangle = -\langle X_3 \rangle -\langle Z_4 X_3 Z_2 \rangle - \langle Z_5X_3 Z_2 \rangle = -1+1+0 = 0, \\
g_{5,2}^{(2)} &= -\sum_{l=1,3,4} \langle Z_l X_5 Z_2 \rangle = -\langle Z_1 X_5Z_2\rangle - \langle Z_3 X_5 Z_2 \rangle - \langle Z_4 X_5 Z_2 \rangle = 0 + 0 +1 = 1, \\
g_{1,4}^{(2)} &= -\sum_{l=2,5} \langle Z_l X_1 Z_4 \rangle = -\langle Z_2 X_1 Z_4 \rangle - \langle Z_5X_1Z_4 \rangle = 1 + 0 = 1, \\ 
g_{3,4}^{(2)} &= -\sum_{l=2,4,5} \langle Z_l X_3 Z_4\rangle = -\langle Z_2 X_3 Z_4 \rangle - \langle X_3 \rangle - \langle Z_5 X_3 Z_4 \rangle = 1-1+0 = 0, \\
g_{5,4}^{(2)} &= -\sum_{l=1,3,4} \langle Z_l X_5 Z_4 \rangle = -\langle Z_1 X_5 Z_4 \rangle - \langle Z_3 X_5 Z_4 \rangle - \langle X_5 \rangle = -1,
\end{align}
\end{subequations}
\end{widetext}
the largest gradients are $g_{5,2}^{(2)}$ and $g_{1,4}^{(2)}$. Since they are equal, we break the tie arbitrarily and chose the pair of qubits $(1,4)$. Thus the state at step $r=2$ is given by 
\begin{multline}
|\psi_2\rangle = e^{i\frac{\pi}{4}Y_1Z_4}|\psi_1\rangle 
= \frac{1}{2}\left[|+-+++\rangle + |--+-+\rangle \right. \\ - \left. |+++-+\rangle - |-++++\rangle \right].
\end{multline}
After the application of the gate we update the records of active and inactive qubits, which now are $[1,2,4]$ and $[3,5]$, respectively. The third set of gradients is given by 
\begin{widetext}
\begin{subequations}
\begin{align}
g_{3,1}^{(3)} &= -\sum_{l=2,4,5} \langle Z_l X_3 Z_1 \rangle = -\langle Z_2 X_3 Z_1 \rangle - \langle Z_4X_3Z_1 \rangle - \langle Z_5 X_3 Z_1 \rangle = 1-1+0 = 0, \\ 
g_{5,1}^{(3)} &= -\sum_{l=1,3,4} \langle Z_l X_5 Z_1 \rangle = - \langle X_5 \langle - \langle Z_3 X_5 Z_1 \rangle - \langle Z_4 X_5 Z_1 \rangle = -1 + 0 -1 = -2, \\
g_{3,2}^{(3)} &= -\sum_{l=2,4,5}\langle Z_l X_3 Z_2 \rangle = -\langle X_3 \rangle - \langle Z_4 X_3 Z_2 \rangle - \langle Z_5 X_3 Z_2 \rangle = -1 + 1 + 0 = 0, \\ 
g_{5,2}^{(3)} &= -\sum_{l=1,3,4} \langle Z_l X_5 Z_2\rangle = -\langle Z_1 X_5 Z_2 \rangle - \langle Z_3 X_5 Z_2 \rangle - \langle Z_4 X_5 Z_2 \rangle = 1 + 0 + 1 = 2, \\
g_{3,4}^{(3)} &= -\sum_{l=2,4,5}\langle Z_l X_3 Z_4 \rangle = -\langle Z_2 X_3 Z_4 \rangle - \langle X_3 \rangle - \langle Z_5 X_3 Z_4 \rangle = 1-1+0 = 0, \\ 
g_{5,4}^{(3)} &= -\sum_{l=1,3,4} \langle Z_{l X_5 Z_4} \rangle = -\langle Z_1 X_5 Z_4 \rangle - \langle Z_3 X_5 Z_4 \rangle - \langle X_5 \rangle = -1 + 0 -1 = -2.
\end{align}
\end{subequations}
\end{widetext}
The largest gradient is $g_{5,2}^{(3)}=2$, the gate is applied at the pair of qubits $(2,5)$, where $2$ is an active qubit and $5$ is inactive. The state at step $r=3$ is given by 
\begin{widetext}
\begin{multline}
|\psi_3 \rangle = e^{i\frac{\pi}{4}Z_2Y_5}|\psi_2\rangle = \frac{1}{2\sqrt{2}}\left[|+-+++\rangle +|++++-\rangle 
+ |--+-+\rangle + |-++--\rangle \right. \\
\left. -|+++-+\rangle - |+-+--\rangle - |-++++\rangle  -|--++-\rangle \right],
\end{multline}
\end{widetext}
with the records of active and inactive qubits updated to $[1,2,4,5]$ and $[3]$, respectively. From this state we can compute the set of gradients of step $r=4$. They are given by 
\begin{widetext}
\begin{subequations}
\begin{align}
g_{1,3}^{(4)} &= -\sum_{l=2,4,5} \langle Z_lX_3Z_1\rangle = -\langle Z_2 X_3 Z_1 \rangle - \langle Z_4X_3Z_1 \rangle - \langle Z_5 X_3 Z_1 \rangle = 1-1+1 = 1, \\
g_{2,3}^{(4)} &= -\sum_{l=2,4,5} \langle Z_l X_3 Z_2 \rangle = -\langle X_3 \rangle - \langle Z_4 X_3 Z_2 \rangle - \langle Z_5 X_3 Z_2\rangle = -1 + 1 - 1 = -1, \\ 
g_{4,3}^{(4)} &= -\sum_{l=2,4,5} \langle Z_l X_3 Z_4 \rangle = -\langle Z_2 X_3 Z_4 \rangle - \langle X_3 \rangle - \langle Z_5 X_3 Z_4 \rangle = +1-1+1 = 1, \\
g_{5,3}^{(4)} &= -\sum_{l=2,4,5} \langle Z_l X_3 Z_5 \rangle = -\langle Z_2 X_3 Z_5 \rangle - \langle Z_4 X_3 Z_5 \rangle - \langle X_3 \rangle = -1 + 1 -1 = -1.
\end{align}
\end{subequations}
\end{widetext}
There are two largest gradients, $g_{1,3}^{(4)}=g_{4,3}^{(4)}=1$. We break the tie arbitrarily and take the pair of qubits $(3,4)$. Thus the state at step $r=4$ is given by 
\begin{widetext}
\begin{multline}
|\psi_4\rangle = e^{i\frac{\pi}{4}Y_3 Z_4}|\psi_3\rangle = \frac{1}{4}\left[ |+-+++\rangle + |+---+\rangle + |++++-\rangle + |++---\rangle + | --+-+\rangle \right. \\ 
+ |---++\rangle + |-++--\rangle + |-+-+-\rangle -|+++-+\rangle - |++-++\rangle - |+-+--\rangle \\
\left. -|+--+-\rangle -|-++++\rangle -|-+--+\rangle - |--++-\rangle -|-----\rangle \right].
\end{multline}
\end{widetext}
To extract the cut found by our algorithm we should write $|\psi_4\rangle$ in the computational basis. In order to do this we use its stabilizers, which are 
\begin{equation}
-XXXXX,\enspace -Z\mathbb{I}\mathbb{I}\mathbb{I}Z, \enspace + \mathbb{I}Z\mathbb{I}\mathbb{I}Z, \enspace - \mathbb{I}\mathbb{I}Z\mathbb{I}Z, \enspace - \mathbb{I}\mathbb{I}\mathbb{I}ZZ \nonumber, 
\end{equation}
which correspond to the state 
\begin{equation}
|\psi_4\rangle = \frac{1}{\sqrt{2}}\left(|10110\rangle - |01001\rangle \right),
\end{equation}
in the computational basis. This state upon a measurement in this basis returns the cut $(\mathcal{A},\overline{\mathcal{A}}) = ([1,3,4],[2,5])$, which is a maximal cut of the graph under consideration. We illustrate this partitioning of the graph by coloring the nodes in $\mathcal{A}$ red and those in $\overline{\mathcal{A}}$ blue, and show the resulting partitioned graph in Fig.~\ref{fig:figure_examples}b. Additionally in Fig.~\ref{fig:figure_examples}c we show the approximation ratio of the states $|\psi_r\rangle$ computed in this section, notice that at $r=4$ we have approximation ratio equal to $1$, indicating the algorithm found a state compose of strings encoding maximal cuts.

\subsection{An example with \texorpdfstring{$N=4$}{\textit{N=4}} nodes}
We consider now the graph with $N=4$ nodes shown in Fig.~\ref{fig:figure_examples}d. Its adjacency matrix is given by 
\begin{equation}
[\omega_{i,j}] = \begin{pmatrix}
0 && 1 && 1 && 0 \\
1 && 0 && 1 && 0 \\
1 && 1 && 0 && 1 \\
0 && 0 && 1 && 0
\end{pmatrix}.
\end{equation}
Let us start the algorithm wiht the state $|\psi_0\rangle = |+-++\rangle$. For this state there are two largest gradients at step $r=1$ given by $g_{2,3}^{(1)} = g_{2,1}^{(1)} = 1$. We break the tie arbitrarily and pick the pair of qubits $(2,3)$. Thus the state at step $r=1$ is given by 
\begin{equation}
|\psi_1\rangle = e^{i\frac{\pi}{4}Y_2Z_3}|\psi_0\rangle = \frac{1}{\sqrt{2}}\left[ |+-++\rangle -|++-+\rangle \right].
\end{equation}
Now, the gradients at step $r=2$ are given by 
\begin{widetext}
\begin{subequations}
\begin{align}
g_{1,2}^{(2)} &= -\sum_{l=2,3} \langle Z_l X_1 Z_2 \rangle = -\langle X_1 \rangle - \langle Z_3 X_1 Z_2 \rangle = -1 + 1 = 0, \\ 
g_{1,3}^{(2)} &= -\sum_{l=2,3} \langle Z_l X_1 Z_3 \rangle = -\langle Z_2 X_1 Z_3 \rangle - \langle X_1 \rangle = 1-1 = 0, \\
g_{4,2}^{(2)} &= -\langle Z_3X_4Z_2 \rangle = 1, \\ 
g_{4,3}^{(2)} &= -\langle X_4 \rangle = -1.
\end{align}
\end{subequations}
\end{widetext}
Thus the largest gradient is $g_{4,2}^{(2)}=1$. We apply the next gate to the pair $(2,4)$ leading to a state at step $r=2$ of the form 
\begin{multline}
|\psi_2\rangle = e^{i\frac{\pi}{4}Z_2Y_4}|\psi_1\rangle = \frac{1}{2}\left[ |+-++\rangle + |+++-\rangle \right. \\ 
\left. - |++-+\rangle - |+---\rangle \right].
\end{multline}
For the next step we find all three gradients $g_{1,2}^{(3)} = g_{1,3}^{(3)} = g_{1,4}^{(3)} = 0$, thus no gate needs to be added in this last step. We verify this by looking at the approximation ratio of the states produced by the algorithm, shown in Fig.~\ref{fig:figure_examples}f, we observe that only after two steps the algorithm reaches approximation ratio of $1$, indicating a maximal cut has been found. In order to extract this cut we write $|\psi_2\rangle$ in the computational basis as 
\begin{equation}
|\psi_2\rangle = \frac{1}{2}\left[|0010\rangle + |1101\rangle -|1010\rangle - |0101\rangle \right],
\end{equation}
notice that the algorithm prepares a state which encodes two distinct maximal cuts for the graph under consideration. One of the form $(\mathcal{A}, \overline{\mathcal{A}}) = ([1,2,4],[3])$ which we illustrate in Fig.~\ref{fig:figure_examples}e, and one of the form $(\mathcal{A}, \overline{\mathcal{A}}) = ([1,3],[2,4])$.

\section{Estimation of the mean approximation ratios for the case of positive weights}
\label{app:estimate_mean_ratio}
In this appendix we present the method used to estimate $\overline{\alpha}$ and $\overline{\alpha}^{\rm r}$ reported in Sec.~\ref{subsec:positive_weight}. Recall that our algorithm solves the problem $N$ times, each time starting from a different position $k\in[1,N]$. As such, for the same problem we might have up to $N$ different $\alpha$\textquotesingle s.

We begin by fixing a threshold value $\alpha_{\rm tr}$ for a given graph ensemble. For $N\in[10,30]$, we count how many initial positions $k$, $\mathrm{Num}(N;\alpha_{\rm tr})$, lead to a solution with $\alpha>\alpha_{\rm tr}$. We repeat this process for all problem instances considered and obtain $\mathbb{E}[\mathrm{Num}(N;\alpha_{\rm tr})]$. At this point, we perform a linear fit to the data $(N, \mathbb{E}[\mathrm{Num}(N;\alpha_{\rm tr})])$ and obtain the slope $\mathcal{M} = \mathcal{M}(\alpha_{\rm tr})$. We then vary the threshold $\alpha_{\rm tr}\in[0.88, 1]$ and repeat the above procedure. Once all the data $(\alpha_{\rm tr}, \mathcal{M}(\alpha_{\rm tr}))$ has been obtained we identify $\overline{\alpha} = \alpha_{\rm tr} |_{ \mathcal{M}(\alpha_{\rm tr})=0}$, the last threshold value before the slope becomes negative. The largest approximation ratio we can guarantee is thus the one for which no initial position $k\le N$ leads to $\alpha = \alpha_{\rm tr}$. To account for fluctuations among instances, the linear fit is done to the data $(N, \mathbb{E}[\mathrm{Num}(N;\alpha_{\rm tr})] - \sigma[\mathrm{Num}(N;\alpha_{\rm tr})])$, with $\sigma[\mathrm{Num}(N;\alpha_{\rm tr})])$ one standard deviation. As reported in the main text, this procedure leads to $\overline{\alpha}=0.9686$ for the case of positive-weighted complete graphs with $N\in[10,30]$ and $100$ instances per $N$.

For the case of randomly chosen initial condition we identify $\overline{\alpha}^{\rm r} = \alpha_{\rm tr} |_{ \mathcal{M}(\alpha_{\rm tr})=0.5}$, that is, the threshold for which at least half of the possible initial conditions will lead to, on average, an approximation ratio equal to the threshold. As reported in the main text, this procedure leads to $\overline{\alpha}^{\rm r}=0.8986$.

\bibliography{clifford_biblio}
\end{document}